\documentclass[fleqn,12pt]{article} 
\usepackage{amsmath,amsfonts,epsfig,mathrsfs,todonotes,wasysym,stmaryrd} 
\usepackage{pict2e,picture} 

\usepackage{color} 
\usepackage{cite} 

\numberwithin{equation}{section}

\newcommand{\pa}{\partial} 
\newcommand{\fff}{\vrule width0.5pt height5pt depth1pt} 
\newcommand{\pp}{{{ =\hskip-3.75pt{\fff}}\hskip3.75pt }} 

\newcommand{\mach}{\vartheta}

\newcommand{\then}{\Rightarrow}

\newcommand{\al}{\alpha}

\newcommand{\de}{\delta} 
\newcommand{\f}{\phi} 
\newcommand{\s}{\sigma} 
\renewcommand{\i}{\iota} 
\newcommand{\vp}{\varphi} 
\newcommand{\La}{\Lambda} 
\newcommand{\Om}{\Omega}

\newcommand{\el}{{\frac{e^L-1}L}} 
\newcommand{\vsp}{{}^{\textstyle\vphantom{X}}}

\newcommand{\Lie }{{\cal L}}

\newcommand{\ber}{\begin{eqnarray}} 
\newcommand{\eer}{\end{eqnarray}} 
\newcommand{\beq}{\begin{equation}} 
\newcommand{\eeq}{\end{equation}} 

\newcommand{\nn}{\nonumber} 
\newcommand{\na}{\nabla} 
\newcommand{\half}{{\textstyle{\frac12}}} 
\newcommand{\ihalf}{{\textstyle{\frac i 2}}}

\newcommand{\bbD}[1]{\mathbb{D}_{#1}} 
\newcommand{\bbDB}[1]{\bar{\mathbb{D}}_{#1}}

\newcommand{\re}[1] {(\ref{#1})}

\newcommand{\ha}{\frac12} 
\newcommand{\gij}{g_{ij}} 
\newcommand{\bij}{b_{ij}} 
\newcommand{\hijk}{H_{ijk}} 
\newcommand{\cl}{\Lie } 

\makeatletter 
\DeclareRobustCommand{\bbnabla}{{\mathpalette\bb@nabla\relax}} 
\newcommand{\bb@nabla}[2]{%
  \begingroup 
  \sbox\z@{$\m@th#1\nabla$}%
  \dimendef\Dht=6 \dimendef\Dwd=8 
  \setlength{\Dwd}{\wd\z@}%
  \setlength{\Dht}{\ht\z@}%
  \begin{picture}(\Dwd,\Dht) 
  \put(0,0){$\m@th#1\nabla$} 
  \put(.35\Dwd,.9\Dht){\line(13,-28){.32\Dht}} 
  \end{picture}%
  \endgroup 
}

\makeatother 

\begin{document} 

\begin{titlepage} 

\begin{flushright} \small
Imperial-TP-2019-CH-01\\
UUITP-51/18\\
YITP-SB-18-31\\
\end{flushright}
\vskip 0.8cm 


\begin{center} 

{\bf \large T-Duality in $(2,1)$ Superspace} 

\vspace{.5cm} 

M.\ Abou-Zeid\footnote{bahom96@gmail.com}, C.\ M.\ Hull\footnote{c.hull@imperial.ac.uk}, U.\
Lindstr\"{o}m${}^{2,}$\footnote{lindoulf@gmail.com} and M.\ Ro\v{c}ek\footnote{martin.rocek@stonybrook.edu}

\vspace{.5cm} $^1${\small \em SUB, Georg-August-Universit\"{a}t G\"{o}ttingen,\\ Platz der G\"{o}ttinger Sieben 1, 37073
G\"{o}ttingen, Germany} 

\vspace{.5cm} $^2${\small \em Theory Group, The Blackett Laboratory,\\ Imperial College London, Prince Consort Road, London
SW7 2AZ, UK} 

\vspace{.5cm} $^3${\small \em Department of Physics and Astronomy, Uppsala University,\\ Box 516, SE-751 20 Uppsala, Sweden}

\vspace{.5cm} {\small \em $^4$C.~N.\ Yang Institute for Theoretical Physics, Stony Brook University,\\ Stony Brook, NY
11794-3840,USA} 

\vspace{1cm} 


\begin{abstract} 
We find the T-duality transformation rules 
for 2-dimensional (2,1) supersymmetric sigma-models
in (2,1) superspace.
 Our results clarify certain aspects of the (2,1) sigma model geometry relevant to the discussion of T-duality. The complexified duality transformations we find   are equivalent to the usual Buscher duality transformations (including an important 
refinement) together with diffeomorphisms.
We  use the gauging of sigma-models in (2,1) superspace, which we review and develop,  finding a manifestly real and geometric expression for the gauged action.
We discuss the obstructions to gauging (2,1) sigma-models, and find that the obstructions to (2,1) T-duality are considerably weaker.
 \end{abstract} 

\end{center} 

\vspace{1cm} 

\end{titlepage} 

\setcounter{footnote}{0} 

\tableofcontents 
\section{Introduction} 

\label{Genintro} 

Supersymmetric nonlinear sigma models with $D$-dimensional target spaces have a rich structure, which makes them good tools
for studying various geometries. 
The target-space geometries are constrained 
by the number of
supersymmetries; in particular, 
there is a direct correspondence between target-space complex structures and world-volume supersymmetries. 
For two dimensional $(p,q)$ supersymmetric models, the relationship between geometry and supersymmetry is particularly rich
\cite{AlvarezGaume:1981hm,GHR,Howe:1985pm,HullWitt,Hull:1985zy,Hull:1986hn,Howe:1987qv,Howe:1988cj,Hull:2016khc,Hull:2017hfa,Hull:2018jkr}. 
The $(2,2)$ models 
of \cite{GHR} have generalised K\"ahler geometry \cite{Hitchin:2004ut,Gualtieri:2003dx}. 
A more general complex geometry with torsion arises for $(2,0)$ supersymmetry \cite{HullWitt}, and for $(2,1)$ supersymmetry
\cite{Hull:1986hn}, while the general geometry for $(p,q)$ supersymmetric models for all $p,q$ was found in
\cite{Hull:1986hn}; see also \cite{Howe:1987qv,Howe:1988cj}. 
The $(2,1)$ supersymmetric models \cite{Hull:1985zy} will be the focus of this paper 
and are relevant for supersymmetric compactifications of ten-dimensional superstring theories as well as for critical superstrings with $(2,1)$ supersymmetry
\cite{Ooguri:1991ie}, which have interesting applications \cite{Kutasov:1996fp,Kutasov:1996zm,Kutasov:1996vh,Green:1987cw}.
Their target space geometries include the generalised K\"{a}hler geometries of the $(2,2)$ models as special cases. 
The reduction of $(2,2)$ models to $(2,1)$ superspace was discussed in ref. \cite{Hull:2012dy}. 
The $(2,1)$ superspace formulation was first given in \cite{DinSeib}.

T-duality relates two-dimensional sigma-models that have different target space geometries but which define the same quantum
field theory; for a review and references, see \cite{Giveon:1994fu}. 
When the target space of a model has  isometry group $U(1)^d$, its T-dual is found by gauging the isometries and adding Lagrange
multiplier terms (plus an important total derivative term) 
\cite{TB,TB2,RV}. Integrating out the Lagrange multipliers 
  constrains the (world-sheet) gauge fields to be trivial and so
gives back the original model, while integrating out the gauge
fields yields the T-dual theory, with the dual geometry given by the Buscher rules \cite{TB}.  Various gaugings in and out of 
superspace have been described in \cite{Hull:1985pq,Kapustin:2006ic,Merrell:2006py,Lindstrom:2007vc,C&B1 , C&B2, Jack:1989ne, gauging, Hull:1993ct, AH1,AH2
,Hull:2006qs, Crichigno:2015pma}.

The starting point for T-duality is the gauging of the sigma model, and extended supersymmetry imposes restrictions on the gauging.
In particular, the isometries must be compatible with the supersymmetries, i.e.\   holomorphic
with respect to all the associated complex structures 
\cite{AlvarezGaume:1983ab,gauging,Hull:1990qh}. For $(2,2)$ supersymmetry, the gauging was 
discussed in \cite{AlvarezGaume:1983ab,gauging, Kapustin:2006ic,Lindstrom:2007vc,Crichigno:2015pma,Merrell:2006py}, while 
the gauging of $(2,1)$ supersymmetric models was given in \cite{ AH1,AH2} for the 
superspace formulation of \cite{DinSeib} 
and in \cite{gauging} for the formulation of \cite{Howe:1987qv,Howe:1988cj}. 

The supersymmetric T-duality transformations have an interesting geometric structure. For sigma models with K\" ahler target
geometry, the T-duality changes the K\" ahler potential by a Legendre transformation \cite{Lindstrom:1983rt,RV}. In general,
duality can change the representation of the supersymmetry \cite{Lindstrom:1983rt}. 
T-duality for $(2,2)$ supersymmetric sigma models has been studied in 
\cite{RV,Grisaru:1997ep,Lindstrom:2007sq,Merrell:2007sr,Crichigno:2011aa}. 

Here we will use the results of \cite{AH1,AH2} to analyse T-duality for $(2,1)$ supersymmetric models in $(2,1)$ superspace
\cite{DinSeib}. 
Adding Lagrange multiplier terms to the gauged theory and integrating out the gauge multiplets gives a dual geometry, with a
$(2,1)$ supersymmetric version of the T-duality transformation rules. 
The supersymmetric gauging involves a complexification of the action of the isometry group, resulting in a T-duality
transformation that 
is a complexification of the usual T-duality rules. 
The
complexified T-duality we find is equivalent to a real Buscher T-duality combined with a diffeomorphism; this is the same mechanism that was previously found    for the $(2,2)$ supersymmetric T-duality   (see \cite{TB,Lindstrom:2007sq}).

For bosonic and $(1,1)$ sigma-models with Wess-Zumino term, there are geometric and topological obstructions to gauging in
general \cite{C&B1,C&B2}. For T-duality, however, the obstructions are considerably milder \cite{Alvarez:1993qi,
Hull:2006qs}. Here we will
extend this discussion to $(2,1)$ models, analysing the obstructions to gauging and T-duality. Moreover, we will interpret
our results for T-duality in terms of generalised moment maps and a generalised K\"{a}hler quotient. 

The paper is organised as follows. In section~\ref{sigmaT}, we first review the general gauged sigma model and the
obstructions to its gauging. We then summarise the formulation   of T-duality of \cite{Hull:2006qs} in terms of a lift to a higher-dimensional sigma
model and show that the obstructions to T-duality are much milder than the obstructions to gauging -- one can T-dualise an ungaugable sigma model. In particular, we recall and emphasize that the Buscher rules are modified when the sigma-model Lagrangian is invariant only up to a total derivative term under the isometry used for the T-duality \cite{Alvarez:1993qi,Hull:2006qs}. In section~\ref{21superspace} we give the superspace description of the $(2,1)$ models. Section~\ref{rigidiso}
discusses the isometries of $(2,1)$ models in superspace. In section~\ref{21YMmultiplet}, we review the superspace
description of the $(2,1)$ Yang-Mills supermultiplet. In section \ref{Gauged21model} we review the results of \cite{AH1,AH2}  on the gauging
of the $(2,1)$ models. We discuss T-duality for the $(2,1)$ sigma models in section~\ref{21duality}, and derive the duality
transformations of the potentials for the $(2,1)$ geometries with torsion. 
We find the duality transformations for the metric and $b$-field, which give a complex version of the Buscher rules.
In section~\ref{sectBusch}, we explain how our complexified T-duality transformations give the real Buscher rules combined with diffeomorphisms, and illustrate this with some examples.
In section~\ref{Obstruct}, we adapt the general results of ref. \cite{Hull:2006qs} to the geometry and T-dualisation of
$(2,1)$ models, including the cases for which there are obstructions to the gauging and for which the standard T-dualisation
procedure fails. Section 10 contains a summary of our results. Some technical details are collected in four appendices. 


\section{The gauged sigma model and T-duality} 
\label{sigmaT} 

\subsection{The gauged bosonic sigma model}

The two-dimensional sigma model with $D$-dimensional target space $M$ is a theory of maps $\f:\Sigma\to M$, where $\Sigma$ is a 2-dimensional
manifold. 
The action is the sum of a kinetic term $S^0_{kin}$ and a Wess-Zumino term $S^0_{WZ}$, 
\beq 
S^0=S_{kin}^0+S_{WZ}^0 \ . 
\label{CBact} 
\eeq 
Given a metric $g$ on $M$ and a  metric $h$ on $\Sigma$, the kinetic term can be written as
\beq 
S_{kin}^0=\ha \int _\Sigma   
*\,  tr (h^{-1} \f ^* g)
\eeq 
where the Hodge dual on $\Sigma$ for the metric   $h$ is denoted by $*$ and $\f ^* g$ is the pull-back of $g$ to $\Sigma$. 
If $x^i$ ($i=1,\ldots D$) are coordinates on $M$ and $\s^a$ are coordinates on $\Sigma$, the map is given locally
by functions $x^i(\s)$ and 
$tr (h^{-1} \f ^* g) = h^{ab} (\f ^* g)_{ab} = h^{ab} g_{ij}\pa _a x^i   \pa _b x^j $, so that 
the Lagrangian 2-form can be written locally as
\beq 
L_{kin}^0=\ha   \gij (x(\s))\, dx^i \wedge *dx^j \ . 
\eeq 
Here and in what follows, the pull-back $\f ^*(dx^i )= \pa _a x^i d \s ^a$ will be written as $dx^i$, 
and it should be clear from the context whether a form on $M$ or its pull-back is intended. 

The Wess-Zumino term 
is constructed using a closed 3-form $H$ on $M$. 
We write
\beq 
S_{WZ}^0= \int _\Gamma \f ^*H \ , 
\label{wzh} 
\eeq 
where $\Gamma$ is any 3-manifold with boundary $\Sigma$. 
 This can be written in terms 
of local coordinates as
\beq 
S_{WZ}^0= 
\frac13\int _\Gamma 
\hijk \, dx^i \wedge dx^j\wedge dx^k \, .
\eeq
Locally, $H$ is given in terms of a 2-form potential $b$ with 
\beq \label{wzb1}
H=db \ , 
\eeq 
and the Wess-Zumino term can be written locally in terms of a 2-form Lagrangian on a patch in $\Sigma$
\beq \label{wzb} 
S_{WZ}^0= \ha
\int _\Sigma
 \bij (x(\s))\, dx^i \wedge dx^j . 
\eeq 

The functional integral involving the Wess-Zumino term (\ref{wzh}) is well-defined 
and independent of the choice of $\Gamma$ 
provided 
$\frac {1}{2\pi} H$ represents an integral cohomology class\footnote{When the third cohomology group $\mathbb H^3$ of M is nontrivial, this leads to a quantisation condition for H; if $\mathbb H^3$ is trivial, then \re{wzb1}  is globally defined and there is no quantisation condition.} on $M$.

The conditions for gauging isometries of this model were derived in \cite{C&B1,C&B2} and will now be briefly
reviewed. Suppose there are $d$ Killing vectors $\xi _K$ ($K=1,\ldots d$) with $\cl_K g=0$, $\cl_K H=0$, where 
$\cl_K$ is the Lie derivative with respect to $\xi _K$. 
The $\xi _K$  generate an isometry group with structure constants 
$f_{KL}{}^M$, 
with 
\beq 
[\cl_K ,\cl _L]=f_{KL}{}^M \cl_M . 
\label{LL} 
\eeq 
Then under the transformations 
\beq \label{trans} 
\de x^i = \lambda ^K \xi _K^i(x) 
\eeq 
with constant parameters $\lambda ^K$, 
the action (\ref{CBact}) changes by 
a surface term if 
$\i _K H$ is exact, so that the equation 
\beq \label{vis} 
\i _K H=du_K 
\eeq 
is satisfied for some (globally defined) 1-forms $u_K$. The $u_K$ are defined by (\ref{vis}) up to the addition of exact
forms. Thus the transformations~\re{trans} are global symmetries provided $\i _K H$ is exact. When this is the case, the
functions 
\beq \label{bis} 
c_{KL} \equiv \i_K u_L 
\eeq 
are globally defined. We note that in the special case in which the $b$-field is invariant,
\beq\label{lbz} 
\cl_K b=0 \, ,\eeq 
we have
\beq
u_K= \i_K b~, 
\eeq
but in general $u_K\neq\i_K b$. 

The gauging of the sigma-model \cite{C&B1,Jack:1989ne,C&B2} consists in promoting the symmetries (\ref{trans}) to local ones, with
parameters that are now functions $\lambda ^K(\s)$, 
by seeking a suitable coupling to connection 1-forms $A^K$ on $\Sigma$ 
transforming as 
\beq \label{ctrans} 
\de A^M =d\lambda^K- f_{KL}{}^M A^K \lambda^L . 
\eeq 
The conditions for gauging to be possible found in \cite{C&B1,C&B2} are that (i)  $\i_K H$ is exact, (ii)  a 1-form $u_K=u_{Ki} dx^i$ satisfying (\ref{vis}) can be
chosen 
that satisfies the equivariance condition 
\beq \label{liev} 
\cl _K u_L= f_{KL}{}^M u_M 
\eeq 
(so that $\i _K H$ represents a trivial equivariant cohomology class \cite{Figueroa-OFarrill:1994vwl}),
and (iii)
\beq \label{ivas} 
\i_K u_L =-\i_L u_K 
\eeq 
so that the globally defined functions (\ref{bis}) are skew, 
\beq \label{cskew} 
c_{KL}=-c_{LK} . 
\eeq 

Defining the covariant derivative of $x^i$ by 
\beq \label{dcis} 
D_a x^i \equiv \pa _a x^i - A_a ^K \xi _K^i 
\eeq 
and the field strength 
\beq \label{Fabc} 
F^M=dA^M - \frac 1 2 f_{KL}{}^M A^K \wedge A^L , 
\eeq 
the gauged action is \cite{C&B1} 
\beq 
S= S_{kin}+ S_{WZW} ~. 
\label{wzhg} 
\eeq 
The gauged metric term is minimally coupled: 
\beq\label{Sggauge} 
S_{kin} = \frac12 \int_\Sigma \gij D x^i \wedge *D x^j ~, 
\eeq 
whereas the gauged Wess-Zumino-Witten term involves a non-minimal term: 
\beq \label{gag3} 
S_{WZW}= \int _\Gamma \left( \frac13 \hijk D x^i \wedge D x^j \wedge D x^k + F^K \wedge u_{Ki} Dx^i \right), 
\eeq 
with $\partial \Gamma = \Sigma$. It was shown in \cite{C&B1,C&B2} that this is closed and locally can be written as
\beq \label{wzb2} 
S_{WZW}= \int _\Sigma 
\left( 
\ha \bij \, dx^i \wedge dx^j +A ^K \wedge u_{K} +\ha 
c_{KL} A^K\wedge A^L 
\right) \ , 
\eeq 
with $u_K=u_{Ki} dx^i $. If the gauge group $G$ acts freely on $M$, then the gauged theory (\ref{wzhg}) gives a quotient
sigma model with target space $M/G$ (the space of gauge orbits) 
on fixing a gauge and 
 eliminating the gauge fields using their equations of motion. 

\subsection{Dualisation}

A general method of dualisation of the  ungauged sigma model (\ref{CBact}) on $(M,g,H)$
is to gauge an isometry group $G$ as above and add the Lagrange multiplier term 
$\int _\Sigma 
F^K \hat x_K $
involving $d$ scalar   fields $\hat x_K
$.
The Lagrange multiplier fields  $\hat x_K $ impose the constraint that the gauge fields $A$ are flat and so pure gauge 
locally, so that (at least locally) one recovers the ungauged model. 
(If $\Sigma$ is simply connected,
e.g. if $\Sigma =S^2$ or 
 $\Sigma =\mathbb{R}^2$,
then $A$ is pure gauge and one recovers precisely  the  ungauged model.) 
Alternatively, fixing the gauge by a suitable constraint on the coordinates $x^i$ and integrating out  the  gauge fields $A$ gives a dual sigma model whose coordinates now include the fields  $\hat x_K
$. This method applies quite generally, including the cases of non-Abelian or non-compact $G$. 

In general, the two dual sigma models are distinct in the quantum theory. However for special cases, the two dual sigma models can define the {\it same} quantum theory, in which case the two dual theories are said to be related by a   T-duality. 
T-dual theories arise for  isometry groups $G$ that are compact and Abelian so that $G=U(1)^d$ with the action of $G$ defining a torus fibration on $M$
  for which    the torus fibres  are the orbits of $G$. There are also further restrictions on the torus fibration; see e.g. \cite{Hull:2006qs}. The classic example is that in which $M$ is a torus $T^d$, with the natural action of $G=U(1)^d$ on the torus.
  
String theory backgrounds require sigma models that define  conformally invariant quantum theories. For a sigma model on  $(M,g,H)$ to define a conformal field theory in general requires the addition of a coupling to a dilaton field $\Phi$ on $M$ through a Fradkin-Tseytlin term, and the T-duality then takes a sigma model on $(M,g,H,\Phi)$
to a dual one $(M',g',H',\Phi')$ on a  manifold $M'$ (which in general is different from $M$), with the two sigma models defining the same conformal field theory. A proof of the quantum equivalence of T-dual CFT's was given in \cite{RV}. 

For applications to T-duality, we focus on the case of Abelian isometries. We derive dual pairs of geometries for general Abelian isometry groups (including non-compact groups or ones that act with fixed points). It is convenient to refer to all of these as T-dualities, although not all lead to full quantum equivalence between dual theories, so not all are proper T-dualities in the strict sense. Our main interest will be in dual pairs that define equivalent quantum theories, but the same formulae apply to the more general class of dual theories.

For Abelian $G$,  
$f_{KL}{}^M=0$, 
so that, assuming $u$ satisfies the equivariance condition \re{liev}, 
\beq 
\cl _K \xi ^L=0, \qquad \cl _K u_L=0, \qquad \cl _K c_{LM}=0 . 
\label{Lxiuc} 
\eeq 
Starting from (\ref{vis}) and  (\ref{Lxiuc}), 
the identity 
\beq \label{iihislv} 
\i_K \i_L H= \cl _ Ku _L - d \i _K u_L 
\eeq 
implies $\i_K \i_L H$ is exact, with 
\beq \label{iihis} 
\i_K \i_L H= -d c_{KL} ~, 
\eeq 
and
\beq \label{iiihis} 
\i_K \i_L \i_M H= 0 \ . 
\eeq 

To dualise the ungauged sigma model (\ref{CBact}) on $(M,g,H)$ with respect to $d$ Abelian isometries, one gauges the
isometries as above and adds the following Lagrange multiplier term involving $d$ scalar Lagrange multiplier fields $\hat x_K
$ \cite{FJ,TB,TB2,Fradkin:1984ai,RV,Giveon:1994mw} 
\beq \label{lmt} 
S_{LM}= \int _\Sigma 
A ^K \wedge d\hat x_K ~. 
\eeq 
This differs from the expression $\int _\Sigma 
F^K \hat x_K $ by a surface term  that is crucial for quantum equivalence \cite{ RV,Giveon:1994mw}.
The $\hat x_K $ impose the constraint that the gauge fields $A$ are flat. For compact $G$, 
the holonomies $e^{i\oint A}$ around non-contractible loops on $\Sigma$
are eliminated by requiring the $\hat x_K $ to be periodic coordinates of a torus $T^d$ so  
that the winding modes of the $\hat x_K $
set the holonomies $e^{i\oint A}$ to the identity.
Then the gauge field is trivial for any $\Sigma$ and $A$ can be absorbed by a gauge transformation, recovering the original ungauged
model. Alternatively, fixing a gauge 
and integrating out the gauge fields $A$ gives the T-dual sigma model. 
In adapted coordinates $x^i=(x^K,y^\mu )$ in which 
$$ \xi ^i_K \frac {\pa } {\pa {x^i}}= \frac {\pa } {\pa {x^K}} , 
$$ 
one can fix the gauge setting the $x^K$ to constants and this gives a dual geometry with coordinates 
$(\hat x_ K,y^\mu )$.

One of the conditions for gauging to be possible was that $c_{KL}=- c_{LK}$.
If one relaxes this constraint, then  (\ref{wzhg}) is no longer gauge invariant,  with
its gauge variation depending on the constants $c_{(KL)}$ and
given by
\beq
\delta S= \int _\Sigma \, c_{(KL)}d\lambda ^ K \wedge
A ^L  ~. 
\eeq 
Remarkably, this variation can then be cancelled by the variation of (\ref{lmt}) by
 requiring that 
 \beq
 \delta \hat x_K = c_{(KL)}\lambda ^L
 \eeq
 so that $\hat x_K $ can be thought of as a compensator field, transforming as a shift under the gauge symmetry.
This was  first
observed  in \cite{Alvarez:1993qi} for the special case of a single isometry and extended to 
the general case in \cite{Hull:2006qs}.
Furthermore, it was shown in \cite{Hull:2006qs} that introducing the fields  $\hat x_K $ through (\ref{lmt}) allows all three conditions for gauging listed above to be relaxed and replaced by one much milder condition. This allows the gauging and T-dualisation of ungaugable sigma models; we next review the construction of \cite{Hull:2006qs}.

\subsection{Duality as a quotient of a higher dimensional space}

It is natural to seek to interpret the Lagrange multiplier fields $\hat x_K $ as $d$ extra coordinates, so that we have a
sigma model with $D+d$ dimensional target space $\hat M$ with coordinates 
$\hat x^\alpha =(x^i, \hat x_K)$, where $\alpha =1,\ldots D+d$. Then the gauged action plus the Lagrange multipler term can
be viewed as a gauge-invariant sigma model on $\hat M$, and this can be compared with the standard form of the gauged sigma
model (\ref{wzhg}) reviewed above. In particular, the terms linear in $A$ in the sum of the Wess-Zumino term (\ref{wzb2}) and
the Lagrange multiplier term (\ref{lmt}) are 
\beq \label{lmtsa} 
S_{LM}= \int _\Sigma
A ^K \wedge (u_K+ d\hat x_K ) , 
\eeq 
which suggests introducing a modified 1-form 
\beq \label{fhrd} 
\hat u_K= u_K+ d\hat x_K 
\eeq 
on $\hat M$. 
If the condition that $u_K$ is a globally defined one-form is dropped, the constraint that $du_K$ is a globally defined closed 2-form suggests interpreting $u_K$ as a connection one-form on a $U(1)^d$ bundle over $M$. If $\hat x_K$ are taken as fibre coordinates, then   $\hat u_K$ can be globally defined one-form on $\hat M$; this is the starting point for the construction of \cite{Hull:2006qs}.

The space $\hat M$ with coordinates 
$\hat x^\alpha =(x^i, \hat x_K)$ 
is then a bundle over $M$ with projection $\pi : \hat M \to M$ which acts as 
$\pi : (x^i, \hat x_K) \mapsto  x^i$. 
A (degenerate) metric $\hat g$ and closed 3-form $\hat H$ can be chosen on $\hat M$ with no $\hat x_K$ components, i.e.\  
\beq \label{abcads} 
\hat g = \pi ^* g, \qquad \hat H = \pi ^* H , 
\eeq 
where $ \pi ^*$ is the pull-back of the projection. The pull-back will often be omitted in what follows, so that the above
conditions will be abbreviated to $\hat g=g, \hat H =H$. Then the only non-vanishing components of $\hat g_{\alpha\beta}$ are
$\gij$, $\partial/\pa \hat x_K$ is a null Killing vector, and the only non-vanishing components of $\hat H
_{\alpha\beta\gamma}$ are $\hijk$. 

We consider the general set-up with $d$ commuting vector fields on $M$ preserving $H$. This implies that there are local
potentials $u_K$ with $\i_K H = d u_K$, but they need not be global 1-forms, and need not satisfy \re{liev} or \re{ivas}.

We lift the Killing vectors $\xi_K$ on $M$ to 
vectors $\hat \xi_K$ on $\hat M$ with 
\beq \label{khatis} 
\hat \xi_K=\xi_K + \Omega _{KL} \frac {\pa} {\pa \hat x_L}~, 
\eeq 
for some $\Omega_{KL}$ to be determined below. 
As the metric $g$ and the torsion 3-form $H$ are both independent of the coordinates $\hat x_K$, the $\hat \xi_K$ are Killing
vectors on $\hat M$: 
\beq 
\hat \cl _K \hat g=0, \qquad \hat \cl _K \hat H=0 . 
\eeq 

As $du= d \hat u$, with $\hat u$ given in \re{fhrd}, it follows that 
\beq \label{Habc} 
\hat \i_K \hat H = d \hat u_K , 
\eeq 
where $\hat \i_K$ denotes the interior product with $\hat \xi_K$. From (\ref{fhrd}), we find 
\beq 
\hat \i_K\hat u_L = \i_K u_L +\Omega _{KL} . 
\eeq 
If we now choose 
\beq 
\label{thetis} 
\Omega _{KL}=- \ha (\i_K u_L +\i_L u_K)~, 
\eeq 
then 
\beq 
\label{hatiu} 
\hat \i_K\hat u_L +\hat \i_L \hat u_K= 0 
\eeq 
and the functions on $\hat M$ defined by 
\beq 
\hat c_{KL}\equiv \hat \i_K\hat u_L 
\eeq 
are found to satisfy 
\beq 
\hat c_{KL}=c_{[KL]} , 
\eeq 
where the functions $c_{KL}$ are defined in (\ref{bis}). 

Next, the Lie derivative of the potentials $\hat u_K$ with respect to $\hat \xi$ is now zero: 
\beq \label{Kabc} 
\hat \cl _K\hat u _L=0 , 
\eeq 
so the $\hat u_K$ are equivariant. Finally, 
if 
\beq 
\label{ijkH} 
\i_K \i_L \i_ M H=0 , 
\eeq 
then the isometry group generated by the 
$\hat \xi_K$ is Abelian, 
\beq 
[ \hat \cl _K, \hat \cl _K]=0 . 
\eeq 
Note that this condition implies that $\hat \i_K\hat \i_L\hat \i_ M\hat H=0$. 

The target space $\hat M$ has dimension $D+d$, where $D$ is the dimension of $M$ and $d$ is the dimension of the 
Abelian gauge group $G$. 
The gauged model on $\hat M$ 
gives, on eliminating the gauge fields, a quotient sigma model with target space given by the space of orbits, $\hat M/G$,
which is also of dimension $D$. The T-dual geometry is given by this quotient space. 

In summary, if we start from a geometry $(M,g,H)$ preserved by $d$ commuting Killing vectors $\xi_K$, 
then on a patch $U$ of $M$ we can find local potentials $u_K$ satisfying $du_K=\i_K H$ 
and lift them to Killing vectors $\hat \xi_K$ and potentials $\hat u_K$ on a patch of $\hat M$. 
If the torsion 3-form $H$ on $M$ satisfies $\i_K \i_L \i_ M H=0$, then there are no 
further local obstructions to gauging the isometries on $\hat M$ 
generated by $ \hat \xi_K$, even when there are local obstructions to gauging the isometries on $ M$ 
generated by $ \xi_K$. For the gauged action on $\hat M$ to be globally defined, 
one needs to specify the bundle over $M$ by giving the transition
functions for the coordinates $\hat x_K$, require that the $ \hat \xi_K$ are globally defined vector fields on $\hat M$ and
also that the $\hat u_K$ are globally defined 1-forms on $\hat M$. 
In the overlaps $U\cap U'$ of patches $U,U'$ on $M$, the
potentials $u_K$ satisfying $du_K=\i_K H$ are related by $ u'_K=u_K+d \alpha _K$ for some transition functions $\alpha_K$, so that the $u_K$ are components of a connection on $M$ with field strength given by $\i_K H$. 
The $\hat x_K$ are then fibre coordinates with 
$\hat u_K = u_K+ d\hat x_K$ globally defined on $\hat M$. 
If the Killing vectors $\xi_K$ can be normalised so that $\frac 1 {2\pi} \i_K H$
 all represent integral cohomology classes, then the bundle can be taken to be a $U(1)^d$ bundle with fibres $(S^1)^d$, while
otherwise it is a line bundle with fibres $\mathbb{R}^d$. Details of the global structure are given in \cite{Hull:2006qs}.
For T-duality, we require that the fibres be circles. Generalisations to cases in which the $\hat \xi_K$, the $\hat u_K$ or
both are only locally defined, or in which $\i_K \i_L \i_ M H\ne 0$, were discussed in
\cite{Hull:2006qs,Hull:2006va,Belov:2007qj,Hull:2008vw}; such T-dualities, when they can be defined, typically lead to non-geometric
backgrounds. 

We remark on an important observation made in \cite{Alvarez:1993qi} for a single isometry and  in \cite{Hull:2006qs} for the general case:  when \re{lbz} is not satisfied, that is, when $\cl_K b\neq 0$ and hence when $u_K\neq\i_K b$, {\em the Buscher rules}
\cite{TB} {\em are modified}. For a single isometry in adapted coordinates $x^i=(x^0, y^\mu)$, $\xi=\pa / \pa x^0$, the 
dual geometry has coordinates $(\hat x^0, y^\mu)$ and the
modified Buscher rules are: 
\beq 
g^D_{\hat0\hat0}=\frac1{g_{00}}~~,~~g^D_{\hat0\mu}=\frac{u_\mu}{g_{00}}~~,~~g^D_{\mu\nu}=g_{\mu\nu}+ 
\frac1{g_{00}}\left(u_\mu u_\nu-g_{0\mu}g_{0\nu}\right)~,\nonumber 
\eeq 
\beq\label{modbusch} 
b^D_{\hat0\mu}=\frac{g_{0\mu}}{g_{00}}~~,~~b^D_{\mu\nu}=b_{\mu\nu}- 
\frac1{g_{00}}\left(u_\mu g_{0\nu}-g_{0\mu}u_{\nu}\right)~. 
\eeq 
The usual Buscher rules are recovered when $u_\mu=b_{0\mu}$. Geometric formulae for the duality transformations for the tensors $g,H$ (without using adapted coordinates) for arbitrary numbers of isometries are given in \cite{Hull:2006qs}.

Finally, the global issues which may arise when T-dualising are dealt with in the standard way. Suppose the coordinate $x^0$
is periodic with $x^0\sim x^0+ 2\pi$, and the metric contains the radii: $g_{00}=R^2$. Here, as throughout the paper, we have set the string tension $T=1$, but to keep track of dimensions, we can introduce it by rescaling the metric $g_{ij}\to Tg_{ij}$, so $g_{00}=TR^2$; then the radius in dimensionless units is 
$\sqrt{T}R$. After we gauge and introduce the dual coordinate $\hat x^0$, we can insure the holonomies of the gauge fields are trivial and hence the model is equivalent to the original ungauged model
by insisting that $\hat x^0$ is periodic with $\hat x^0\sim \hat x^0+2\pi$. Consider the functional integral given by
\beq 
\int \, [Dx^i D\hat x^0 D A ^K] \, e^{i(T\, S+S_{LM})} ,
\label{functI} 
\eeq 
where $S$ is the gauged sigma model action (\ref{wzhg}), and $S_{LM}$ is the Lagrange multiplier term (\ref{lmt}).  Then (\ref{functI}) 
is invariant under large gauge transformations for compact world-sheets $\Sigma$ 
of arbitrary topology, and using the Buscher rules we have
\beq
T\hat R^2 =g^D_{\hat0\hat0}=\frac1{g_{00}}=\frac1{TR^2}~~\then~~ \hat R=\frac1{TR}~.
\eeq

The analysis of the geometry, gauging and T-duality given in this section for bosonic sigma models readily extends to (1,1)
supersymmetric sigma models formulated in (1,1) superspace: the geometry of the gauging is just as in the bosonic case. 
For such (1,1) models to have $(2,1)$ supersymmetry requires the existence of a complex structure 
with certain restrictions on the geometry. For the gauging to be possible with manifest $(2,1)$ supersymmetry requires the
Killing vectors to be holomorphic. The geometry of the gauged $(2,1)$ sigma models and their application to T-duality will be
analysed in the following sections. 

\section{The $(2,1)$ sigma model in superspace} 
\label{21superspace} 

The $(2,1)$ 
superspace is parametrised by two Bose coordinates $\sigma^{\pp} ,\sigma^{=}$, a complex Fermi chiral spinor coordinate
$\theta^{+},\bar\theta^+$, and a single real Fermi 
coordinate $\theta^{-}$ of the opposite chirality. It is natural to define the complex conjugate 
left-handed spinor derivatives 
\beq 
D_+=\frac{\partial}{\partial \theta^+} +i 
\bar\theta^{+}\frac{\partial}{\partial \sigma^{\pp}} \ \ ,\ \ 
\bar D_+=\frac{\partial}{\partial \bar \theta^+\vsp} +i 
\theta^+\frac{\partial}{\partial \sigma^{\pp}} , 
\label{superD1} 
\eeq 
as well as a real right-handed spinor derivative 
\beq 
D_{-}=\frac{\partial}{\partial \theta^{-}} +i 
\theta^{-}\frac{\partial}{\partial \sigma^{=}}. 
\eeq 
These spinor derivatives satisfy the algebra 
\beq 
D_+^2 = 0~,~~~\bar{D}_+^2= 0 ~,~~~ D_-^2 = i\partial_{=} ~,~~~ \left\{ D_+ , \bar{D}_+ \right\} = 2i\partial_{\pp} . 
\eeq 

We denote by $M$ the $D$ real dimensional target space manifold of the sigma model and pick local coordinates $x^i$, $i=1,
\ldots D$ in which the metric and torsion potential are $g_{ij}$ and $b_{ij}$. It was shown in \cite{HullWitt,GHR,Hull:1985pq,Hull:1986hn} that
invariance of the (1,1) supersymmetric sigma model action under a second (right-handed) chiral supersymmetry requires that

{\em (i)} $D$ is even 

{\em (ii)} $M$ admits a complex structure\footnote{Supersymmetric models with almost complex structures were considered in
\cite{deWit:1988fk,Delius:1989nc}; they obey a modified supersymmetry algebra and are not considered here.}  $J^i{}_j$ 

{\em (iii)} the metric is
hermitian with respect to the complex structure and 

{\em (iv)} the
complex structure $J^i{}_j$ is covariantly constant with respect to the connection $\nabla ^+=\nabla + \frac 1 2 g^{-1}H$ with torsion $\frac 1 2 g^{-1}H$.

We assume that these conditions are satisfied so that the sigma model has
$(2,1)$ supersymmetry. We choose a complex coordinate system $z^\alpha$, 
$\bar{z}^{\bar{\beta}}=(z^\beta )^*$, $(\alpha , 
\bar{\beta}=1\ldots \frac{1}{2}D)$ in which the line element is $ds^2 
=2g_{\alpha \bar{\beta}}dz^{\alpha} d \bar{z}^{\bar{\beta}}$ and the complex structure is constant and diagonal, 
\beq 
J^i{}_j = i\left( \begin{array}{cc} \delta^{\beta}{}_\alpha & 0 \\ 
0 & -\delta^{\bar{\beta}}{}_{\bar{\alpha}} \end{array} \right) . 
\eeq 
The supersymmetric sigma 
model can then be formulated in $(2,1)$ superspace in terms of scalar 
superfields 
$\vp^\alpha$, $\bar \vp^{\bar{\alpha}} = ( \vp^\alpha )^*$, which are constrained to satisfy the chirality conditions 
\beq 
\bar{D}_{+}\vp^\alpha =0 \ \ , \ \ D_+ 
\bar{\vp}^{\bar{\alpha}} =0 . \label{chiral} 
\eeq 
The lowest components of the superfields 
$\vp^\alpha |_{\theta =0} =z^\alpha$ are the bosonic complex coordinates 
of $M$. The most general renormalizable and Lorentz invariant $(2,1)$ 
superspace action written in terms of chiral scalar superfields is  \cite{DinSeib} 
\beq 
S = S_1 + S_2 , 
\label{full21action} 
\eeq 
where 
\beq 
S_1 =i\int d^2 \sigma d\theta^+ d\bar\theta^+ d\theta^- \left( k_\alpha 
D_- \vp^\alpha -\bar{k}_{\bar{\alpha}}D_- 
\bar{\vp}^{\bar{\alpha}} \right) 
\label{21action} 
\eeq 
and 
\beq 
S_2 =i\int d^2 \sigma d\theta^+ d\theta^- F(\vp ) + \mbox{complex conjugate} . 
\label{21superpotential} 
\eeq 
Here F is a holomorphic section, as it is defined only up to the addition of a complex constant. Since F depends only on chiral superfields, the integration in \re{21superpotential}  is over $\theta^+$ only (and not over $\bar \theta^+$).  The term  $S_2$ is the analogue of the F-term in four dimensional supersymmetric field theories. In particular, this term can spontaneously break supersymmetry, it is not generated in sigma model perturbation theory if it is not present at tree level, and it is subject to a nonrenormalisation theorem, so that it is not corrected from its tree level value (up to possible wave-function renormalisations).

The $(2,1)$ sigma model geometry is sometimes referred to as strong K\"{a}hler with torsion (or SKT for short). 
It is determined locally by the complex vector field $k_\alpha(z,\bar{z})$ with complex conjugate
\beq 
(k_\alpha) ^* = \bar{k}_{\bar \alpha} . 
\label{realcond1} 
\eeq 
The metric, torsion potential and torsion are given by 
\ber 
g_{\alpha \bar{\beta}} & = & \partial_{\alpha} 
\bar{k}_{\bar{\beta}}+ 
\bar \partial_{\bar{\beta}}k_{\alpha} \nn \\ 
 b'_{\alpha \bar{\beta}} 
& = & 
\partial_{\alpha}\bar{k}_{\bar{\beta}}-\bar \partial_{\bar{\beta}} 
k_{\alpha} \nn \\ H_{\alpha \beta \bar{\gamma}} & = & \frac{1}{2} 
\bar \partial_{\bar{\gamma}} \left( \partial_{\alpha}k_{\beta} 
-\partial_{\beta} 
k_{\alpha}\right) ~, 
\label{geometry} 
\eer 
where $ b'$ is the torsion potential in a gauge where it is purely $(1,1)$. 
If the torsion $H=0$, the manifold $M$ is K\"{a}hler with $k_\alpha = 
\half \frac{\partial}{\partial z^{\alpha}}K(z ,\bar{z})$ 
where $K(z ,\bar{z})$ is the K\"{a}hler potential, and the 
$(2,1)$ supersymmetric model actually has $(2,2)$ supersymmetry, while for $H 
\neq 0$, $M$ is a hermitian manifold with torsion of the type 
introduced in \cite{HullWitt,GHR}. 

The torsion potential $b_{ij}$ is only defined up to an 
antisymmetric tensor gauge transformation of the form 
\beq 
\delta b_{ij} =\partial_{[i} \lambda_{j]} . 
\label{btransf} 
\eeq 
The (1,1)-form potential 
\beq 
 b'\equiv b'_{\alpha \bar\beta}d\bar{z}^{\bar\beta}\!\wedge\! d{z}^\alpha=(\bar
k_{\bar\beta,\alpha}-k_{\alpha,\bar\beta})\,d\bar{z}^{\bar\beta}\!\wedge\! d{z}^\alpha\ , 
\label{b11form} 
\eeq 
can be transformed to a (2,0)+(0,2) form by a gauge transformation
\beq
b'\to b'+ d(\bar k_{\bar\beta}d\bar{z}^{\bar\beta}+k_\beta
d{z}^\beta) =b^{(0,2)}+b^{(2,0)}
\eeq
where
\beq
\label{b02} 
b^{(2,0)}=k_{\beta,\alpha}d{z}^{\alpha}\!\wedge\!
d{z}^{\beta}, \qquad 
b^{(0,2)}=\bar k_{\bar\beta,\bar\alpha}d\bar{z}^{\bar\alpha}\!\wedge\! d\bar{z}^{\bar\beta}~.
\eeq

The geometry (\ref{geometry}) is preserved by the transformation 
\beq\label{symm} 
\delta k_\alpha = \tau_\alpha 
\eeq 
provided $\tau_\alpha$ satisfies 
\beq\label{trans_315} 
\bar{\partial}_{\bar{\beta}}\tau_\alpha = 
i\partial_\alpha \bar{\partial}_{\bar{\beta}}\chi 
\eeq 
for some arbitrary real $\chi$. 
This implies that $\tau$ is of the form 
\beq\label{rhogen} 
\tau_\alpha = i\partial_\alpha \chi +\mach_\alpha \ \ , \ \ 
\bar{\partial}_{\bar{\beta}}\mach_{\alpha} = 0 
\eeq 
for some holomorphic $\mach_\alpha$. The symmetry (\ref{symm}) is 
the analogue of the generalised K\"{a}hler transformation 
discussed in \cite{GHR}. It leaves the metric and torsion invariant, but 
changes 
$b_{ij}$ by an 
antisymmetric tensor gauge transformation of the form (\ref{btransf}). 

\section{Isometries in the $(2,1)$ sigma model} 
\label{rigidiso} 
For the application to T-duality discussed in the following sections, we shall be interested in Abelian groups of isometries. For
completeness, however, we discuss the general case of non-Abelian isometry groups.

Let $G$ be a group of isometries of $M$ generated by Killing vector fields $\xi^i_K$ that preserve the metric and 3-form $H$,  $\cl_K g=0$, $\cl_K H=0$, and satisfy the algebra (\ref{LL}).
This symmetry will be consistent with (2,1) supersymmetry if  
\beq 
(\Lie_K J)^i{}_j = 0 \label{dJ=0} ~.
\eeq 
This allows us to write the symmetry
of the $(2,1)$ 
supersymmetric model in (2,1) superspace as 
\beq 
\delta \vp^i = \lambda^K \xi_{K}^{i}(\vp ) \label{iso21} ~.
\eeq 
The  constraint (\ref{dJ=0})
is the condition that
the $\xi^i_K$ are 
holomorphic Killing vectors\footnote{ A discussion of sigma models with non-holomorphic isometries can be found in ref. \cite{Hull:1990qh}.} 
with respect to the complex structure $J_{ij}$, giving 
\beq \label{xiholo} 
\partial_\alpha \bar{\xi}^{\bar{\beta}}_K = 0 
\eeq 
in complex coordinates.
If the torsion vanishes, then $M$ is K\"{a}hler, and the K\"ahler 2-form 
$\omega$ (with components $\omega_{ij}\equiv g_{ik}J^k{}_j$) is closed. For every holomorphic 
Killing vector $\xi^i_K$, the 1-form with components $\omega_{ij} 
\xi^j_K$ is closed and locally there are functions $X_K$ such that $\omega_{ij} 
\xi^j_K =\partial_i X_K$; in complex coordinates, this   equation becomes 
$\xi_{K\alpha} = 
i\partial_\alpha X_K$. 
The functions $X_K$ are
sometimes called Killing potentials and 
  play a central role in gauging the supersymmetric 
sigma-models without 
torsion \cite{BaggWitt,Hull:1985pq}. When $X_K$ are globally defined  equivariant functions (i.e. $\Lie_K X_M=0$), they are referred to as moment maps and the gauging implements the K\" ahler quotient construction.

When the torsion does not vanish, this generalises 
straightforwardly \cite{gauging}. 
The locally defined 1-form $u_K$ satisfies
$\i _K H=du_K $.
If, in addition, (\ref{dJ=0}) holds, then the 1-form with 
components $\nu_i \equiv \omega_{ij}(\xi^j_K +u^j_K)$ satisfies 
$\partial_{[\alpha}\nu_{\beta ]}=0$, so that 
there are generalised Killing potentials such 
that \cite{gauging} 
\beq 
\xi_{\alpha K} + u_{\alpha K} = \partial_\alpha Y_K +i\partial_\alpha X_K . 
\label{defX} 
\eeq 
The $X_K$ and $Y_K$ are locally defined functions on $M$; $Y_K$ simply reflects the ambiguity in the definition of 
$u_K$ in \re{vis}, 
and locally the $\partial_\alpha Y_K $ term can be absorbed into the definition of $u_K$.

Under the rigid symmetries (\ref{iso21}), the variation of the 
action in (\ref{21action}) is 
\beq 
\delta S_1 =i \lambda^K \int d^2 \sigma d\theta^+ d\theta^-  \left(( \Lie_K k_\alpha) D_- \vp^\alpha 
-(\Lie_K \bar{k}_{\bar{\alpha}})D_- 
\bar{\vp}^{\bar{\alpha}} \right) , 
\label{varLrig} 
\eeq 
where the Lie derivative of $k_\alpha$ is 
\beq 
\Lie_K k_\alpha = \xi^{\beta}_K \partial_\beta k_\alpha + 
\bar{\xi}^{\bar{\beta}}_K \bar \partial_{\bar{\beta}}k_\alpha 
+k_\beta \partial_\alpha \xi^{\beta}_K . \label{Liek} 
\eeq 

The variation of the superpotential term (\ref{21superpotential}) in the action is 
\beq 
\delta S_2 = i \lambda^K \int d^2 \sigma d\theta^+ d\theta^- \Lie_K F(\vp ) + \mbox{complex conjugate} , 
\label{varLFI} 
\eeq 
so it will be left invariant by the isometries provided the holomorphic function $F(\vp )$ is invariant up to constants,
i.e.\  if the equations 
\beq 
\Lie_K F = e_K \label{LF=0} 
\eeq 
are satisfied for some complex constants $e_K$. 

In general, the isometry symmetries 
will not leave the potential $k_\alpha$ invariant, but will change it by 
a gauge transformation of the form (\ref{symm})-(\ref{rhogen}), 
so that the action (\ref{21action}) is unchanged. 
The geometry 
and Killing potentials then determine the quantity $\Lie_K k_{\alpha }$ 
appearing in the variation~(\ref{varLrig}) to take the form 
\beq 
\Lie_K k_\alpha = i\partial_\alpha \chi_K +\mach_{K\alpha} , 
\label{ichitheta} 
\eeq 
for some real functions $\chi_K$ and holomorphic 1-forms 
$\mach_{K\alpha}$, 
\beq 
\bar \partial_{\bar{\beta}}\mach_{K\alpha}=0 . 
\label{tholo} 
\eeq 
In ref. \cite{AH1}, the following explicit expressions for $\chi$ and $\mach$ were found: 
\beq 
\chi_K = X_K +i\left( \bar{\xi}^{\bar{\beta}}_K 
\bar{k}_{\bar{\beta}} 
-\xi^{\beta}_K k_\beta \right) 
\label{chi} 
\eeq 
\beq 
\mach_{K\alpha} = 2\xi^{\gamma}_K \partial_{[\gamma}k_{\alpha ]} 
+\xi_{\alpha K} -i\partial_\alpha X_K . 
\label{theta} 
\eeq 
Using (\ref{xiholo}), (\ref{defX}), and \re{Liek}, it is straightforward to check that (\ref{chi}) and (\ref{theta}) satisfy
(\ref{ichitheta}) and (\ref{tholo}) respectively. 
It follows that the action of the 
Lie bracket algebra on the vector potential $k_\alpha$ 
reduces to 
\beq 
[ \Lie_K ,\Lie_L ] k_\alpha = f_{KL}{}^M \Lie_M k_\alpha , 
\label{algk} 
\eeq 
as it must (cf.~(\ref{LL})). 
The obstructions to gauging of the supersymmetric sigma model (without superpotential) were analysed in \cite{AH1,AH2}
following \cite{C&B1, C&B2,gauging}. It was found that, in order for the gauging to be possible, the following two conditions
must hold: 
\ber
(i)\qquad&&\xi^{\alpha}_{(I}\vartheta_{J)\alpha }  =0 \nn\\[2mm]
(ii)\qquad&&\Lie_K X_L = f_{KL}{}^M X_M ~.
\label{Xequiv} 
\eer
Condition {\em (ii)} is   the statement that the generalised Killing  potentials must be equivariant. If they are also globally defined, then they are sometimes referred to as generalised moment maps.

Observe that,  together with the relation~(\ref{defX}), the expression~(\ref{theta}) for $\mach_{J\alpha}$ implies 
\beq 
\xi^{\alpha}_{(I}\vartheta_{J)\alpha }  = \xi^{\alpha}_{(I}u_{J)\alpha } 
\label{c(ab)u} 
\eeq 
(as can be seen by contracting $\mach_{J\alpha}$ with $\xi^\alpha_I$ and symmetrising with respect to $I$ and $J$), so that
condition {\em (i)} above is equivalent to 
\beq 
c_{(IJ)}  =0 
\label{c(IJ)=0} , 
\eeq 
where the functions  $c_{IJ}$ were defined in~(\ref{bis}); compare eq.~(\ref{cskew}). 

For the gauging of the superpotential term term (\ref{21superpotential}) to be possible, it is necessary that the constants
$e_K$ defined in (\ref{LF=0}) vanish, so that the holomorphic function $F(\vp )$ is invariant under the isometry symmetries,
\beq 
\Lie_K F = 0 \label{LF=02} . 
\eeq 

Consider the case of gauging one  isometry that acts in adapted coordinates $(\vp^0,
\vp^\mu)$ as a shift in 
$i(\vp^0-\bar \vp^0)$,  so that $\vp^0 \to \vp^0 +i \lambda$ , $\bar\vp^0 \to\bar \vp^0 -i \lambda$.
The Killing vector $\xi $ then has components $(i,-i,0,\dots )$, with
\beq 
\xi^i\frac\partial{\partial x^i}=i\left(\frac \partial{\partial \vp^0}- 
\frac \partial{\partial \bar \vp^0}\right) . 
\label{adaptedcoor} 
\eeq 
Then the condition (\ref{c(IJ)=0}) implies that 
\beq 
c =\xi^0\vartheta_0=0 ~~\Rightarrow~~ \vartheta_0=0 ~,
\label{cthxi0} 
\eeq 
which, combined with (\ref{theta}) implies that 
\beq 
\pa_0 X=\pa_{\bar 0} X=g_{0\bar 0} . 
\label{dXdXg} 
\eeq 

\section{The $(2,1)$ gauge multiplet and gauge symmetries} 
\label{21YMmultiplet} 

We now promote the 
isometries (\ref{iso21}) to local ones in which the 
constant parameters $\lambda^K$ are replaced by $(2,1)$ superfields $\La^K$, 
\beq 
\delta \vp^\alpha = \La^K \xi^{\alpha}_K \ \ , \ \ \delta 
\bar{\vp}^{\bar{\alpha}} = \bar{\La}^K 
\bar{\xi}^{\bar{\alpha}}_{K} . 
\label{iso} 
\eeq 
These transformations preserve the chirality constraints (3.5) only if the $\Lambda^K$ are chiral,
\beq 
\bar{D}_{+}\La^K = 0 \ \ , \ \ D_+ \bar{\La}^K = 0 . 
\label{Lachir} 
\eeq 
Under a finite transformation, 
\beq 
\vp \rightarrow \vp ' = e^{L_{\La \cdot \xi}}\vp \ \ , \ \ 
\bar{\vp} \rightarrow \bar{\vp} ' 
= e^{L_{\bar{\La} \cdot \bar{\xi}}}\bar{\vp} , 
\label{isoloc} 
\eeq 
where 
\beq 
L_{\La \cdot \xi}   \equiv \La^K \xi^{\alpha}_{K}\frac{\partial}{\partial 
\vp^{\alpha}} 
\eeq 
is the generator of the 
infinitesimal diffeomorphism with parameter $\La \cdot \xi$.

The $(2,1)$ super Yang-Mills multiplet is given in $(2,1)$ superspace 
by a set of Lie-algebra valued super-connections ${\cal A}_{(2,1)}=(A_{+},\bar A_{+},A_{-}, 
A_{\pp},A_{=})$, with $A_{\bullet}=A_{\bullet}^KT_K$, where the Lie algebra generators $T_K$ are 
hermitian and satisfy the algebra $[T_K,T_L]=if_{KL}^{~~~M}T_M$. 
These connections can be used to define gauge covariant 
derivatives $\nabla_\bullet\equiv D_\bullet-iA_\bullet$, which are constrained by the conditions: 
\beq
\left\{ \nabla_+ , \bar{\nabla}_+ \right\}= 2i\nabla_\pp 
~,~~~ \left\{ \nabla_{-} , \nabla_{-} \right\} = 2i\nabla_{=} ~,~~~ 
\left\{ \nabla_+ , \nabla_-\right\} = \bar W ~,~~~ \left\{ \bar{\nabla}_{+} , \nabla_{-} \right\} = W ~,~~~ 
\label{21YMdef} 
\eeq 
as well as $\na_+^2=\bar\na_+^2=0$. 
The remaining relations among the derivatives follow from these conditions and the Bianchi identities, e.g.\  
\beq \label{56}
\left[\na_+,\na_\pp\right]=\frac1{2i}\left[\na_+,\left\{\na_+,\bar\na_+ \right\}\right]=0~,~~~ 
\bar\na_+ W= \left[\bar\na_+,\left\{\bar\na_+,\na_- \right\}\right]=0 ~,~~~ 
\eeq 
\beq \label{57}
\left[\na_+,\na_=\right]=\frac1{2i}\left[\na_+,\left\{\na_-,\na_- \right\}\right]=i\left[\na_-,\left\{\na_-,\na_+
\right\}\right] 
=i\na_-\bar W~, 
\eeq 
\ber \label{58}
\left[\na_-,\na_\pp\right]\!\!&=&\!\!\frac1{2i}\left[\na_-,\left\{\na_+,\bar\na_+\right\}\right]= 
\frac{i}2\left[\na_+,\left\{\bar\na_+,\na_- \right\}\right]+\frac{i}2\left[\bar\na_+,\left\{\na_+,\na_- \right\}\right] 
\nn\\[2mm] 
\!\!&=&\!\!\frac{i}2\left(\na_+W+\bar\na_+\bar W\right)~, 
\eer 
\beq \label{59}
\left[\na_\pp,\na_=\right]=\frac1{2i}\left[\na_\pp,\left\{\na_-,\na_-\right\}\right]= 
i\left[\na_-,\left[\na_-,\na_\pp\right]\right] 
=-\frac12\na_-\left(\na_+W+\bar\na_+\bar W\right)~.~~~ 
\eeq 
The conditions \re{21YMdef} were introduced in \cite{AH1,AH2}. Their consequences \re{56}-\re{59} correct statements in \cite{AH1,AH2}.

The constraints (\ref{21YMdef}) can be solved to give all connections in 
terms of a scalar prepotential $V$ and the spinorial connection $A_{-}$. 
In the chiral representation, the spinorial derivatives that appear in the 
algebra (\ref{21YMdef}) are given by 
\ber 
\bar{\nabla}_{+} =\bar{D}_{+}~,~~~ \nabla_+ = e^{-V} D_+ e^{V} ~,~~~ 
\na_-\equiv D_--iA_-~, 
\label{21nabla-} 
\eer 
where $V=\bar V$ is hermitian, and the spinor connection $A_-$ is hermitian up to a similarity transformation because we are
in chiral representation\footnote{Real representations are reviewed in Appendix \ref{realrep}.}: $\bar
A_-=e^V(A_-+iD_-)e^{-V}$. We then find 
\ber 
\nabla_{\pp }\!\!&\equiv&\!\! -\frac{i}2 \left\{\bar D_+, e^{-V} D_+ e^{V}\right\} 
=\partial_{\pp} -\frac{i}{2 } \bar{D}_+ D_+ V 
+O(V^2)~,\nn\\[2mm] 
\na_=\!\!&\equiv&\!\!-i\na_-^2=\pa_=-(D_-A_-)+iA_-^2~, 
\label{nablapp21} 
\eer 
so that 
\beq 
A_{\pp } = \frac12 \bar{D}_+ D_+ V +O(V^2)~,~~A_==-iD_-A_-+O(A_-^2)~. 
\label{App21} 
\eeq 
The field strengths are obtained from~(\ref{21YMdef}) and~(\ref{21nabla-}), 
\beq 
\bar W \equiv \left\{ e^{-V} D_+ e^{V} , D_- -iA_ -\right\} =-iD_+(A_--iD_-V)+O(VA_-,V^2)~, 
\label{Wbarmin21} 
\eeq 
\beq 
W \equiv \left\{\bar D_+ , D_- -iA_ -\right\}=-i\bar D_+ A_-~. 
\label{Wmin21} 
\eeq 
Again, these are not complex conjugates because we are in a chiral representation.
Note that if instead we used the anti-chiral representation, we would have
 $\bar W= -iD_+A_-$, and $W=\{e^V\bar D_+ e^{-V}, D_--iA_-\}$.

We now turn to the gauge 
transformations of the $(2,1)$ Yang-Mills supermultiplet. Under a finite gauge transformation, the hermitian superfield 
prepotential $V$ transforms as 
\beq 
e^{V} \rightarrow e^{V '}=e^{i\bar\La} e^{V} e^{-i\La} ~.
\label{eV'} 
\eeq 
For infinitesimal $\Lambda$, this yields 
\beq 
\delta V = i(\bar\La-\La) - \frac{i}{2} \left[ V 
, \La + \bar{\La} \right] +O(V^2)~. 
\label{transfV21} 
\eeq 
In chiral representation, the superconnection $A_-$ transforms as 
\beq 
\na_-'=e^{i\La}\na_-e^{-i\La}~~\then~~\delta A_- = \na_- \La ~. 
\label{transfA} 
\eeq 
The antichiral representation would be reached from this by a similarity transformation with $e^V$, giving the antichiral representation covariant derivative
\beq 
\na_\bullet^{(AC)}=e^V \na_\bullet e^{-V}~.
\eeq 
The spinor covariant derivative $\na_-$ is real in the sense that after taking the adjoint 
one is in the antichiral representation: $\bar\na_-=e^V \na_-e^{-V}$. 
In particular, 
\beq 
\delta \bar A_- = \na_-\bar\La~. 
\label{transfAbar} 
\eeq 

For fields in a linear representation of the gauge group, the Lie algebra generators act in that representation. For the   superfields $\vp,\bar\vp$, the symmetry is realised non-linearly, with the Lie algebra element $T_K$ generating the transformation
$\vp\to \vp + \lambda^K\xi_K(\vp)$.

Covariant derivatives can act on different representations of the group. This action is encoded in the matrix used to 
represent the generators of the Lie algebra; they can also act nonlinearly on the superfields $\vp,\bar\vp$. In this case,
the covariant derivative  uses this non-linear realisation: 
\beq 
\na_\bullet \vp^\al\equiv D_\bullet\vp^\al-A^K_\bullet \xi^\al_K~,~~~ 
\bar\na_\bullet \bar\vp^{\bar\al}\equiv\bar D_\bullet\bar\vp^{\bar\al}-\bar A^K_\bullet\bar\xi^{\bar\al}_K~. 
\eeq 

The scalar superfields 
$\vp , \bar{\vp}$ transform under the local isometry 
symmetries as in (\ref{iso}). Following \cite{Hull:1985pq}, we define the chiral-representation version of $\bar\vp$ as 
\beq 
\tilde{\vp} = e^{L_{V\cdot \bar{\xi}}}\bar{\vp} ~, 
\label{deffitilde} 
\eeq 
where 
\beq 
L_{V\cdot \bar{\xi}} \equiv iV^K \bar{\xi}^{\bar{\alpha}}_K 
\frac{\partial}{\partial \bar{\vp}^{\bar{\alpha}} }~ . 
\label{LV} 
\eeq 
Then the superfields $\vp ,\tilde{\vp}$ satisfy the covariant 
chirality constraints 
\beq 
\bar{\nabla}_+ \vp^\alpha = 0 \ \ , \ \ \nabla_+ 
\tilde{\vp}^{\bar{\alpha}} = 0 ~, 
\eeq 
and transform under the isometry symmetries as 
\beq 
\delta \vp^\alpha = \La^K \xi^\alpha_K \ \ , \ \ \delta 
\tilde{\vp}^{\bar{\alpha}} =\La^K 
\bar {\xi}^{\bar{\alpha}}_{K} (\tilde{\vp})~ . 
\label{dfifitilde} 
\eeq 
Here $\bar{\xi}^{\bar{\alpha}}_{K} 
(\tilde{\vp})$ is obtained from $\bar{\xi}^{\bar{\alpha}}_{K} 
(\bar{\vp})$ by replacing $\bar{\vp}$ with $\tilde{\vp}$.
Note that the transformation of $\tilde{\vp}$ involves the parameter 
$\La$, 
while that for $\bar{\vp}$ involves $\bar{\La}$. The 
covariant derivatives of $\tilde{\vp}$ are in chiral representation, and hence are given by: 
\beq 
\nabla_{\bullet} \tilde{\vp}^{\bar{\alpha}} = D_{\bullet} 
\tilde{\vp}^{\bar{\alpha}} -A_{\bullet}^{K} 
\bar{\xi}^{\bar{\alpha}}_{K} 
(\tilde{\vp})~ . 
\label{nablafitilde} 
\eeq

When gauging one translational isometry, we again choose adapted coordinates $\vp^0$ in which the Killing vector has
components $(i,-i,0,\dots)$ and acts as in (\ref{adaptedcoor}). Then the above relations simplify: the only fields that
transform are 
\ber 
\delta \vp^0& \!\!= \!\!&i\La~,\qquad\qquad\qquad 
\delta \bar \vp^{\bar 0} = -i\bar\La ~,\qquad\qquad~~~ 
\delta\tilde\vp^{\bar 0} = -i\La~, \nn \\ 
\delta V & \!\!= \!\! & i\left( \bar\La -\La \right)~,\qquad~\, 
\delta A_- = D_-\La~, 
\label{inftransfadapt} 
\eer 
and minimal coupling is simply given by 
\beq 
\tilde \vp^{\bar 0}=\bar\vp^{\bar 0}+V~. 
\eeq 
As explained above, since the transformation $\delta A_- = D_-\La$ involves 
the chiral parameter $\La$, it is necessarily complex, and $A_-$ is not  real; however, the combination 
\beq\label{realA} 
A_--\ihalf D_-V 
\eeq 
has the real transformation 
\beq\label{realAt} 
\delta(A_--\ihalf D_-V)=\half D_-(\La+\bar\La)~, 
\eeq 
and is real -- see Appendix \ref{realrep} for details. 

\section{The gauged $(2,1)$ sigma model} 
\label{Gauged21model} 

The gauged $(2,1)$ sigma model in superspace was studied in \cite{AH1,AH2} and the full nonpolynomial gauged action was
constructed in \cite{AH2} using the methods of ref. \cite{Hull:1985pq}. We now briefly summarise the main results of the
analysis, referring the reader to these papers for the derivations and further details of the construction. 

Under the infinitesimal rigid transformations (\ref{iso21}), the variation of the $(2,1)$ full superspace Lagrangian 
\beq 
L_1 =i\left( k_\alpha D_- \vp^\alpha -\bar{k}_{\bar{\alpha}} 
D_- \bar{\vp}^{\bar{\alpha}} \right) \label{L21} 
\eeq 
is given by (\ref{varLrig})
\beq 
\delta L_1 =i\lambda^K\!\left((\Lie_Kk_\al) D_-\vp^\al-(\Lie_K\bar k_{\bar\al})D_-\bar\vp^{\bar\al}\right) .
\label{varL1local} 
\eeq 
Invariance of the action requires (\ref{ichitheta}): 
\beq 
\Lie_K k_\alpha = i\partial_\alpha \chi_K +\mach_{\alpha K} \ , 
\label{dk=rho} 
\eeq 
with $\chi_K$ a real function and $\mach_{K\alpha}$ a holomorphic 1-form 
which were shown in ref. \cite{AH1} to take the explicit forms (\ref{chi}) 
and (\ref{theta}) respectively. The variation of the superpotential term (\ref{21superpotential}) is given in (\ref{varLFI}),
which vanishes provided the function $F$ 
satisfies (\ref{LF=02}), i.e.\  if it is invariant under the rigid isometries (\ref{iso21}). 

Now consider promoting the rigid isometries to local symmetries (\ref{iso}). The variation of the $(2,1)$ superpotential term
(\ref{21superpotential}) is given by 
\beq 
\delta S_2 = i \int d^2 \sigma d\theta^+ d\theta^- \La^K \Lie_K F(\vp ) + \mbox{complex conjugate} 
\label{varLFIlocal} 
\eeq 
and this will vanish provided the function $F(\vp )$ is itself invariant under the local isometries, i.e.\  (\ref{LF=02})
holds for such isometries; in the following we will assume that this is the case and concentrate on the gauging of the full
superspace term (\ref{21action}) in the action. 

The main result of \cite{AH2} is that the $(2,1)$ superspace action \re{21action}
can be gauged provided the geometric condition \re{Xequiv}  
holds, in which case the gauge invariant superspace Lagrangian for the gauged $(2,1)$ 
 sigma model is (to all orders in the gauge coupling, which we have absorbed into the gauge fields) 
\beq 
L_{1g} = 
\left[ i\left( k_\alpha D_- \vp^\alpha -\bar k_{\bar{\alpha}}D_- 
\tilde{\vp}^{\bar{\alpha}}\right) - A_{-}^K X_K \right]\! (\vp , 
\tilde{\vp}) 
- \frac{e^L -1}{L} V^K \bar{\mach}_{\bar{\alpha}K}D_- 
\bar{\vp}^{\bar{\alpha}} ~. 
\label{Lfin} 
\eeq 
The operator $L\equiv iV^K \bar{\xi}^{\bar{\alpha}}_K 
\frac{\partial}{\partial \bar{\vp}^{\bar{\alpha}} }$ is the one defined in (\ref{LV}), 
and the expression $\frac{1}{L} (e^L -1)$ 
in the Lagrangian can be defined by its Taylor series expansion in $L$ 
or equivalently by $\int_0^1 dt \,e^{tL}$. 
The gauge invariance of the action obtained from integrating the Lagrangian~(\ref{Lfin}) over superspace is proven in
Appendix \ref{actionprops} for the case of a single isometry. 

The full gauged sigma model $(2,1)$ superspace action is then 
\beq 
S_{tot}= \! \int \! d^2 \sigma\, d^2 \theta^+ d\theta^- L_{1g}\,+\left(\int\! d^2 \sigma\, d \theta^+ d\theta^- F(\vp) \,+\,
c.c.\right) 
\eeq 
for an invariant superpotential $F(\vp)$: $\Lie_K F=0$. 

This form of the gauged action was given in \cite{AH2}, but is not immediately comparable to the more geometric gauged action
given for the bosonic model in \re{Sggauge},\re{wzb2}. As shown in Appendix \ref{actionprops}, using the relations 
\beq\label{urels} 
i(k_{\al,\bar0}+\bar k_{\bar0,\al})=u_\al-iX_{,\al }~,~~~\vartheta_\al=i(k_{\al,0}-k_{0,\al})-u_\al 
\eeq 
we can rewrite the gauged Lagrangian as (for the case of a single isometry -- the general case is similar): 
\ber\label{Lgeo} 
L_{1g} &\!\!= \!\!&i\left( k_\alpha D_- 
\vp^\alpha -\bar k_{\bar{\alpha}}D_- 
\bar{\vp}^{\bar{\alpha}} \right) \! (\vp , \bar\vp) 
- (A_--\ihalf   D_-V) X(\vp ,\tilde{\vp})\nn\\[1mm] 
&&+\,V\frac{e^L -1}{L} 
\left[(u_\al -\ihalf   X_{,\al})D_-\vp^\al +(\bar u_{\bar\al}+\ihalf   X_{,\bar\al})D_-\bar\vp^{\bar\al}\right]~. 
\eer 
Because this Lagrangian is geometric, some properties that are hard to see in \re{Lfin} are more transparent in this form. For
example, the hermiticity of the action follows directly: the combination $A_--\ihalf   D_-V$ \re{realA} is real, with the
real transformation \re{realAt} 
\beq 
\delta(A_--\ihalf   D_-V)= \half D_-(\La+\bar\La)~. 
\eeq 
Since \re{Lxiuc} and \re{Xequiv} imply that  $u_\al$ and $X$ are invariant under (rigid) gauge transformations, we have, for any
real function $f$, $f(\cl_{\xi+\bar\xi})X=0$. In particular, this implies 
\beq 
f(L)X=f([L-\bar L]+\bar L)X = f([iV\cl_{\bar\xi+\xi}]+\bar L)X ~~\then~~f(L)X=f(\bar L)X ~,~~ 
\eeq 
where $\bar L =-iV\xi^\al\frac\pa{\pa\vp^\al}$ and the holomorphy of $\xi$ implies $[L,\bar L]=0$. Similarly 
\beq 
f(L)i [u_\al D_-\vp^\al-\bar u_{\bar\al} D_-\bar\vp^{\bar\al}]= 
f(\bar L)i [u_\al D_-\vp^\al-\bar u_{\bar\al} D_-\bar\vp^{\bar\al}]~. 
\eeq 
The hermiticity of the action then follows.

\section{T-duality of $(2,1)$ supersymmetric theories} 
\label{21duality} 
\subsection{Generalities} 

The generalisation of T-duality \cite{KY,WS,FJ,TB,TB2,Fradkin:1984ai,Giveon:1994mw,Giveon:1994fu} to conformally invariant
sigma models which admit isometries, and its explicit form in $(2,2)$ superspace, were elucidated in ref. \cite{RV}. In this
section we generalise the construction to the superspace formulation of the $(2,1)$ supersymmetric sigma models reviewed
above. The general procedure which defines the dual pairs locally is the same as in refs.
\cite{WS,FJ,TB,TB2,Fradkin:1984ai,RV,Giveon:1994mw,Giveon:1994fu}. First, gauge the sigma model isometries and add a Lagrange
multiplier term constraining the gauge multiplet to be flat. Second, eliminate the gauge fields by solving their field
equations. Classically, this ensures the equivalence of the dual models, modulo global issues that arise in the case of
compact isometries if the gauge fields have nontrivial holonomies along noncontractible loops. However, as already explained
at the end of section \ref{sigmaT}, these issues are taken care of by giving the Lagrange multipliers suitable
periodicities and adding a total derivative term, so that the holonomies are constrained to be trivial \cite{RV}; we will assume
that this can also be done in the $(2,1)$ supersymmetric case at hand. Quantum mechanically, the duality in the $(2,2)$ case
receives corrections from the Jacobian obtained upon integrating out the gauge fields, which at one loop leads to a simple
shift of the dilaton \cite{TB}. 

For an Abelian gauging, the field strengths $W,\bar W$ in chiral representation  given in (\ref{Wbarmin21},\ref{Wmin21}) are: 
\beq 
\bar W ^K=-iD_+(A_-^K-iD_-V^K)~,~~~ 
W^K=-i\bar D_+ A_-^K~. 
\label{WWB} 
\eeq 
The condition that the gauge multiplet is pure gauge can be imposed by constraining $W,\bar W$ to vanish by adding to the
Lagrangian~\re{Lgeo} a term 
\beq 
L_{\Theta} = -\Psi_{K-} W^K -\bar \Psi_{K-} \bar W ^K~. 
\label{Lagrange2Kbis} 
\eeq 
To this we add a total derivative term, which is important for constraining the holonomies of the flat connections correctly, to obtain
\beq\label{Lagrange2K}
L_{\Theta} = -\left( \Theta _K+ \bar{\Theta }_K\right) A_-^K -iD_- \bar{\Theta}_K V ^K, 
\eeq 
where $\Theta=-i\bar D_+\Psi_-,\bar\Theta=-iD_+\bar\Psi_-$ 
are chiral (respectively antichiral) Lagrange multiplier superfields. 
The full action to consider is then 
\beq 
S_{1g} +S_{\Theta}, \qquad S_{\Theta}\equiv  \! \int \! d^2 \sigma d^2 \theta^+ d\theta^- \! L_{\Theta} . 
\label{LfingagK} 
\eeq 

Integrating out the Lagrange multipliers $\Theta$ or $\Psi_- $ gives 
\ber 
W^K=0, \qquad \bar W^K=0 ~, 
\label{wiso} 
\eer 
which implies that $V$ and $A_-$ are pure gauge  (with the boundary terms constraining the holonomies): 
\beq 
A_-^K= D_- \La  ^K~~,~~~~ V^K= i \left(\bar \La ^K- \La ^K \right) . 
\label{21puregaugeAV1K} 
\eeq 
The term $S_\Theta$ then vanishes, and we recover the original sigma model with action
(\ref{full21action})-(\ref{21superpotential}). Alternatively, integrating out the gauge fields gives the T-dual theory. 

In the special case of one isometry, we can choose local complex coordinates $ \{z^\al , \bar z^{\bar \al} \} = \{ z^0 ; \bar
z^{\bar 0} ; z^\mu ; \bar z^{\bar \mu}\}$, with $\mu , \bar \mu = 1,\ldots \frac{D}{2} -1$, such that the isometry acts by a
translation leaving $(z^0+\bar z^0)$ invariant. Moreover we can use a diffeomorphism combined with a $b$ field gauge
transformation to 
arrange for the metric and $b$ field to depend only on $(z^0+\bar z^0)$ and on the set of coordinates $ \{ z^\mu ; \bar
z^{\bar \mu}\}$, but to be independent of $i(z^0-\bar z^0)$; however, if we use a geometric formulation, there is no need to
do so. 
The indices $\mu, \bar \mu$ now run over the `spectator' coordinates transverse to $z^0, \bar z^{\bar 0}$. As reviewed in
the previous section, the requirement of $(2,1)$ supersymmetry restricts the admissible isometries to those that act
holomorphically on chiral superfields.

\subsection{Computations} 

\label{21dualcasegeneric} 

Recall that the geometry constrains $\Lie_K k_\al$ to take the form 
$\Lie_K k_\alpha = i\partial_\alpha \chi_K +\mach_{K\alpha} $
with 
\ber\nn 
&&\bar \partial_{\bar{\beta}}\mach_{K\alpha}=0 \\[1mm]\nn 
&&\chi_K = X_K +i\left( \bar{\xi}^{\bar{\beta}}_K 
\bar{k}_{\bar{\beta}} 
-\xi^{\beta}_K k_\beta \right)\\[1mm] 
&&\mach_{K\alpha} = 2\xi^{\gamma}_K \partial_{[\gamma}k_{\alpha ]} 
+\xi_{\alpha K} -i\partial_\alpha X_K 
\label{tjosan} 
\eer 
(as follows from \re{ichitheta}, \re{tholo}, \re{chi} and \re{theta}) . 

We can perform the T-duality starting from either of the two forms of the gauged Lagrangian,  \re{Lfin} or \re{Lgeo}; for completeness, we consider both. 

\subsubsection{T-duality from the gauged Lagrangian \re{Lfin}}\label{21Lfindual} 

In addition to the $X_K$, we define new potentials 
\beq\label{Zdef} 
 Z_K= X_K+2i \bar{\xi}^{\bar{\beta}}_Kk_{\bar\beta}~. 
\eeq 
Since the Abelian isometries are independent, we focus on only one of them for the sake of clarity and henceforth drop the
index $K$. Splitting the indices as $\alpha =(0,\mu),$ and using coordinates adapted to the isometry $\xi^0=i$ and
$\xi^\mu=0$, from (\ref{tjosan}), we find 
\ber\label{KG}\nn 
X&=&-(k_0+\bar k_{\bar 0})+\chi\\[1mm]\nn 
 Z& \equiv &X+2\bar k_{\bar 0}~~=~~-k_0+\bar k_{\bar 0}+\chi\\[1mm]\nn 
\vartheta_0 & = & -i(g_{0\bar 0} + \partial_0 X) = 0 \nn \\ \vartheta_\mu & = & -i \left[ ( \partial_\mu k_0 -\partial_0
k_\mu ) + g_{\mu \bar 0}+ \partial_\mu X \right]~;
\eer 
it can be checked that $\vartheta_\alpha$ is holomorphic, $\pa_{\bar\beta}\vartheta_\alpha=0$. This is the set-up in the
case where the obstructions to gauging vanish (cf.\ eq.\ (\ref{Xequiv})). However, as we shall see in
section~\ref{Obstruct}, this is not the most general situation in which T-duality is possible. 

Now consider the general gauged Lagrangian in \re{Lfin} with a single translational isometry.  In adapted  coordinates
we have 
\beq\label{adap2} 
\tilde \vp^{\bar\mu}=\bar\vp^{\bar\mu}~, ~~\tilde \vp^{\bar 0}=\bar\vp^{\bar 0}+V ~,
\eeq 
and the gauge invariance of the Lagrangian is shown in Appendix \ref{actionprops}. 

For later purposes, we may rewrite \re{Lfin} as 
\ber\label{Lag2}\nn 
L_g&=&i(k_\alpha D_-\vp^\alpha-\bar k_{\bar \alpha}D_-\bar\vp^{\bar \alpha})- 
X(A_--\ihalf   D_-V)-\ihalf   ZD_-V\\[1mm] 
&& 
-\left(\el\bar{\mach}_{\bar{\mu}}(\bar\vp)\right) VD_-\bar\vp^{\bar\mu}~. 
\eer 
Here  $k_\alpha,\bar k_{\bar\alpha}, X,  Z$ are all functions of 
$\vp,\tilde\vp\equiv e^L\bar\vp$, whereas $\bar\mach$ is a function of $\bar\vp$. 

We add to the general Lagrangian \re{Lag2} the invariant $L_{\Theta}$  \re{Lagrange2K} and consider $L_T=L_g+L_\Theta$ where $\Theta$ and $\bar\Theta$ are chiral and antichiral superfield Lagrange multipliers. As discussed above, integrating out $\Theta$ and $\bar\Theta$ sets the field strengths \re{WWB} to zero (modulo boundary terms):
\beq
\bar W \equiv  - \bar{D}_+ A_- =0, \qquad W \equiv D_+ \left( A_- -iD_- V\right)  =0 \, , 
\eeq
so that the gauge multiplet is pure gauge: $A_{-}  =  D_- \La$, $V = i(\bar\La - \La)$ \re{21puregaugeAV1K}.
Shifting $\vp \to \vp +i\La$, we recover the original ungauged action \re{21action}.

To find the dual action, we integrate out the gauge fields instead. Specialising once again to one isometry, the variation of
$V$ gives an expression for 
$A_-$; however, since $A_-$ enters as a Lagrange multiplier, it drops out of the final 
Lagrangian.\footnote{For completeness, we give the calculation of $A_-$ in Appendix \ref{derAm}.} 
The variation of $A_-$ implies 
\ber\label{Aeq} 
X(\vp,\tilde\vp)+ \Theta + \bar{\Theta}=0~, 
\label{Xthth} 
\eer 
which should be solved for $V=V(\Theta+\bar\Theta,\vp,\bar\vp)$. We can eliminate all dependence on $\vp^0,\bar\vp^{\bar0}$
by choosing the gauge  $\vp^0 =0$; 
in this gauge, $\tilde\vp^{\bar0}=V$, and 
\beq\label{eLsimp} 
V\el\bar\mach(\bar\vp)\equiv \int_0^1dt\, e^{tL}V\bar\vartheta_{\bar\mu}(\bar\vp^{\bar0},\bar\vp^{\bar\nu})~~\to~~ 
\int_0^1dt\, V\bar\vartheta_{\bar\mu}(tV,\bar\vp^{\bar\nu})~. 
\eeq 
Then (\ref{Xthth}) implies 
\ber\nn 
\frac {dX}{d\Theta}=\frac {dX}{d\bar\Theta}=-1 &\iff& V,_\Theta=-\frac1{X,_{\bar 0}}~,~~ V,_{\bar\Theta}=-\frac1{X,_{\bar
0}}~,\\[1mm] 
\frac {dX}{d\vp^\mu}=\frac {dX}{d\bar\vp^{\bar\mu}}= 0 &\iff& V,_\mu 
=-\frac {X,_\mu}{X,_{\bar 0}}~,~~V,_{\bar\mu} 
=-\frac {X,_{\bar\mu}}{X,_{\bar 0}} ~. 
\eer 

We also need the following expression for $D_-V$, which we find by differentiating \re{Aeq} and using the last equation in \re{tjosan} (which
gives  $X,_0=-g_{0\bar0}$): 
\ber\label{DV2} 
D_-V=\frac 1{g_{0\bar 0}}\left(D_-(\Theta+\bar\Theta)+X_{,\mu}D_-\vp^\mu+X_{,\bar\mu}D_-\bar\vp^{\bar\mu}\right)~.
\eer 
Using these results, we now evaluate $L_g+L_\Theta$ from \re{Lag2} and \re{Lagrange2K} to find the dual Lagrangian: 
\ber\nn\label{DLag} 
L^{(D)}&=&i\left(\half \left[V-\frac { Z} {g_{0\bar 0}}\right]D_-\Theta-\half \left[V+\frac { Z} {g_{0\bar
0}}\right]D_-\bar\Theta 
+\left[k_\mu-\half\frac { ZX_{,\mu}} {g_{0\bar 0}}\right]D_-\vp^\mu\right.\\[1mm] 
&&~~~~\left.-\left[\bar k_{\bar\mu}-i\int_0^1dt\,V\bar\vartheta_{\bar\mu}(tV,\bar\vp^{\bar\nu})+\half\frac { ZX_{,\bar\mu}}
{g_{0\bar 0}}\right]D_-\bar\vp^{\bar\mu}\right) , 
\eer 
where $V(\vp,\bar\vp,\Theta+\bar\Theta)$ is found by solving \re{Aeq}. 

\subsubsection{T-duality from the geometric form  \re{Lgeo}  of the gauged Lagrangian.}
Using the equivalent geometric form of the gauged Lagrangian \re{Lgeo} together with the Lagrange multiplier term \re{Lagrange2K} instead, the dual
Lagrangian reads 
\ber\nn\label{LgeoD} 
\hat L^{(D)} &\!\!= \!\!&i\left( \half V D_-\Theta -\half V D_-\bar\Theta+ [k_\mu-iV\frac{e^L -1}{L}(u_\mu-\ihalf  
X_{,\mu})]D_-\vp^\mu\right.\\ 
&&\left.-[\bar k_{\bar\mu}+iV\frac{e^L -1}{L} 
(\bar u_{\bar\mu}+\ihalf   X_{,\bar\mu})D_-\bar\vp^{\bar\mu}]\right)~. 
\eer 

\subsection{The dual geometry } 

From \re{DLag}, we can identify the components of the dual vector potential $k^D$ as follows 
\ber\nn\label{dualks2} 
&&k^D_\Theta=\half \left[V-\frac { Z } {g_{0\bar 0}}\right]\\[1mm]\nn 
&&\bar k^D_{\bar\Theta}=\half \left[V+\frac { Z} {g_{0\bar 0}}\right]\\[1mm]\nn 
&&k^D_{\mu}=\left[k_\mu-\half\frac { ZX_{,\mu}} {g_{0\bar 0}}\right]\\[1mm] 
&&\bar k^D_{\bar\mu}=\left[\bar k_{\bar\mu}- 
i\int_0^1dt\,V\bar\vartheta_{\bar\mu}(tV,\bar\vp^{\bar\nu})+\half\frac { Z X_{,\bar\mu}} {g_{0\bar 0}}\right] . 
\eer 
Note that $\bar k^D_{\bar\mu}$ differs from the complex conjugate of $k^D_{\mu}$ by a 
 a complex transformation of the form \re{symm}-\re{rhogen}, so $k^D$ differs from  a real vector by such a transformation.
Likewise, from \re{LgeoD} we read off 
\ber\nn\label{dualks2r} 
&&\hat k^D_\Theta=\half V\\[1mm]\nn 
&&\bar{\hat k}^D_{\bar\Theta}=\half V\\[1mm]\nn 
&&\hat k^D_{\mu}= [k_\mu-iV\frac{e^L -1}{L}(u_\mu-\ihalf   X_{,\mu})]\\[1mm] 
&&\bar{\hat k}^D_{\bar\mu}=[\bar k_{\bar\mu}+iV\frac{e^L -1}{L} 
(\bar u_{\bar\mu}+\ihalf   X_{,\bar\mu})]~. 
\eer 
Here $\bar{\hat k}^D_{\bar\mu}$ is the complex conjugate of $\hat k^D_{\mu}$ so $\hat k^D$ is a real vector.
Formulae \re{dualks2} and \re{dualks2r} only differ by terms that do not affect the metric and $b$ field. 
 We can  calculate the components of the dual metric $g^D$ and of the dual $b$-field $b^D$. Using \re{geometry}, we find 
\ber\nn 
&&g^D_{\Theta\bar\Theta}=\frac 1{g_{0\bar 0}}\\[1mm]\nn 
&&g^D_{\mu\bar\Theta}=\frac 1{g_{0\bar 0}}[b_{\mu 0}+i\vartheta_\mu ]=\frac {-iu_\mu} {g_{0\bar 0}}\\[1mm]\nn 
&&g^D_{\bar\mu\Theta}=\frac 1{g_{0\bar 0}}[b_{\bar\mu \bar 0}-i\bar\vartheta_{\bar\mu} ] =\frac {i\bar u_\mu} {g_{0\bar
0}}\\[1mm]\nn 
&&g^D_{\mu\bar\mu}=g_{\mu\bar\mu}-\frac 1{g_{0\bar 0}}\left[ {g_{\mu \bar 0}g_{\bar\mu 0}-}(b_{\mu 0}+i\vartheta_\mu)(
b_{\bar\mu \bar 0}-i\bar\vartheta_{\bar\mu} ) \right] 
=g_{\mu\bar\mu}-\frac 1{g_{0\bar 0}}\left[ {g_{\mu \bar 0}g_{\bar\mu 0}-}u_\mu\bar u_{\bar\mu}  \right]\\ 
~\label{metdubu} 
\eer 
and 
\ber\nn 
&&b^D_{\Theta\mu}=\frac {g_{\bar 0 \mu}}{g_{0\bar 0}}\\[1mm]\nn 
&&b^D_{\bar\Theta\bar\mu}=\frac {g_{0 \bar\mu}}{g_{0\bar 0}}\\[1mm]\nn 
&&b^D_{\mu\nu}=b_{\mu\nu}-\frac {2}{g_{0\bar 0}}g_{\bar 0[\mu}(b_{\nu ] 0}+i\vartheta_{\nu ]})=b_{\mu\nu}+\frac {2i}{g_{0\bar
0}}g_{\bar 0[\mu}u_{\nu ] }\\[1mm] 
&&b^D_{\bar\mu\bar\nu}=b_{\bar\mu\bar\nu}-\frac {2}{g_{0\bar 0}}g_{0[\bar\mu}(b_{\bar\nu ] \bar 0}-i\vartheta_{\bar\nu
]})=b_{\bar\mu\bar\nu}-\frac {2i}{g_{0\bar 0}}g_{ 0[\bar\mu}\bar u_{\bar\nu ] }~~~. 
\label{Bdubu} 
\eer 
In the case of $N$ Abelian isometries the expressions for the dual geometry  involve $N\times N$ matrices replacing some
entries, for example $g_{0\bar 0}\to (g+b)_{mn}$ as in the bosonic case \cite{Giveon:1994mw}. 

\section{Comparison to the Buscher rules} 
\label{sectBusch} 

The results \re{metdubu},\re{Bdubu} for the (2,1) duality transformations are similar but not identical to the Buscher transformations in the modified form \re{modbusch}. In the   Buscher duality \re{modbusch}, a coordinate $x^0$ is replaced by a dual coordinate $\hat x^0$ (e.g. if 
$x^0$ is a coordinate on a circle of radius $R$, $\hat x^0$ is a coordinate on the dual circle of radius $2\pi /R T$, again reinstating the string tension to keep track of dimensions), whereas in the (2,1) duality transformations, a {\it complex} coordinate $z^0=\vp ^0|_{\theta =0} $ is replaced by a dual {\it complex} coordinate $\hat z^0=\Theta |_{\theta =0} $. 
This arises because 
the $(2,1)$ gauging involves the action of the {\it complexification} of the isometry group. 
As was explained in 
\cite{TB,Lindstrom:2007sq} for the $(2,2)$ case, the complex duality transformation consists of a T-duality and a diffeomorphism: it gives a T-duality transformation of the imaginary part of the coordinate $z ^0$ and a coordinate transformation of the real part.
Writing $z ^0=  y^0+i x^0$, $\hat z^0= \hat y^0+i \hat x^0$, the 
(2,1) duality transformation consists of a T-duality transformation in which 
the coordinate $x^0$ is replaced by a dual coordinate $\hat x^0$ (so that if 
$x^0$ is a coordinate on a circle of radius $R$, $\hat x^0$ is a coordinate on the dual circle of radius $2\pi /R T$), while $\hat y^0$ is related to $  y^0$ by a coordinate transformation
(so that if 
$y^0$ is a coordinate on a circle of radius $R$, $\hat y^0$ is a different coordinate on the same circle of radius $R$).

To see this, we start from the constraint
 \re{Xthth}: 
\ber 
X(\vp ^0 + \tilde \vp ^ {\bar 0}, \vp^\mu ,    \tilde\vp ^ {\bar \mu})+ \Theta + \bar{\Theta}=0~. 
\label{X8} 
\eer 
Then setting $\theta =0$ and choosing the Wess-Zumino gauge in which 
$V|_{\theta =0}=0$, this implies 
\ber 
X(z ^0+  \bar z ^ {\bar 0}, z^\mu ,   \bar z ^ {\bar \mu})+ \hat z^0 + \bar{\hat z}^{\bar 0}=0~,
\eer 
which gives
\ber 
X(2y ^0 , z^\mu ,   \bar z ^ {\bar \mu})+ 2\hat y^0  =0~. 
\eer 
The solution of this gives
$\hat y^0$ as a function of $y ^0 , z^\mu ,   \bar z ^ {\bar \mu}$, so that the complex duality transformation gives the coordinate transformation
\beq
y ^0\to 
\hat y^0(y ^0 , z^\mu ,   \bar z ^ {\bar \mu})
\eeq
together with the T-duality transformation replacing $x^0$ with the dual coordinate $\hat x^0$.

This can also be understood by comparing our (2,1) superspace analysis with the corresponding computation in  (1,1)  
superspace which gives the Buscher duality  \re{modbusch}: the equivalence of the two calculations is guaranteed, and so will relate the (2,1) duality transformations to the Buscher ones.  The explicit
calculations are carried out in Appendix \ref{redux}.
We now illustrate this discussion with two simple and instructive examples.

\subsection{T-duality on the complex plane}

Our first  simple example is
the complex plane dualised with respect to the isometry given by a rotation about the origin\footnote{This example is interesting in that
it shows that the flat plane, which, when regarded as a string background, has no winding modes, is formally dual to a singular
geometry with no normalizable (radial) momentum modes and only winding modes.}
\beq
 z\to e^{ i\lambda} z
\eeq
for real $\lambda$.
The adapted coordinates are $\vp=\ln z$, transforming under the isometry by an imaginary shift 
$\vp\to \vp + i \lambda$. The metric is given by
\beq\label{cpxds8}
ds^2=dz d\bar z = e^{\vp+\bar\vp}d\vp d\bar\vp~,
\eeq
for which the potential can be taken to be
\beq
k_0=\bar k_{\bar0}=\half e^{\vp+\bar\vp}~.
\eeq
In this case, the Lagrangian is invariant, and $\vartheta=\chi=0$, so \re{KG} gives
\beq
X=-e^{\vp+\bar\vp}~.
\eeq
On gauging, this becomes $X=-e^{\vp+\bar\vp+V}$
and, on choosing the gauge  $\vp=0$,  this reduces to
\beq
X=-e^V~,
\eeq
so that \re{X8} implies
\beq
V=\ln(\Theta + \bar{\Theta})~.
\eeq
Then using \re{dualks2r}, we have
\beq
\hat k^D_\Theta=\bar{\hat k}^D_{\bar\Theta}=\half V=\half\ln(\Theta + \bar{\Theta})~,
\eeq
and 
\beq\label{cpxD8}
g_{\Theta\bar\Theta}=\frac1{\Theta+\bar\Theta}~~\Rightarrow~~d\hat s^2=\frac1{\Theta+\bar\Theta}{\,d\Theta \,d\bar\Theta}~.
\eeq
In real coordinates $\vp=y+ix$, the line element \re{cpxds8} is
\beq
ds^2=e^{2y}(dx^2+dy^2)~,
\eeq
and the isometry is generated by $\pa_x$. Dualizing gives
\beq\label{rD8}
d\hat s^2=e^{2y}dy^2+e^{-2y}d\hat x^2~.
\eeq
To compare this line element to \re{cpxD8}, we write $\Theta=\hat y +i\hat x$, and use the coordinate transformation \re{X8}:
\beq
e^{2y}=(\Theta + \bar{\Theta})=:2 \hat y~.
\eeq
Then \re{cpxD8} becomes:
\beq
 d\hat s^2=\frac1{2\hat y}(d\hat y^2+d\hat x^2)=e^{-2y}(e^{4y}dy^2+d\hat x^2)~,
 \eeq
 which does indeed match \re{rD8}.
 
\subsection{T-duality on a torus}

Consider a flat  torus $S^1 \times S^1$ parametrised by a single complex coordinate $z$ and let $\vp$ be the $(2,1)$ superfield such that $\vp \vert \equiv z$. For simplicity, we consider the case of a single holomorphic isometry and we suppress all spectator fields.
We take the flat metric on the torus to be
\beq
ds^2=R^2 (dx^2 +dy^2) = R^2 dzd\bar z
\eeq
with
\beq
z= y+ix~.
\eeq
The coordinate $x$ that we are dualizing is scaled so that its periodicity is
\beq
x\sim x + 2\pi~, 
\eeq
so it parameterises a circle of  circumference  $2\pi R$ and  $R$; the coordinate $y$ can have any periodicity:
\beq
y\sim y+ \tau~,
\eeq
so the circumference of the corresponding circle is $\tau R$.

We consider the $(2,1)$ sigma model whose target space has the above geometry (with zero $b$-field). This is defined by the potential
\beq
k_\vp= \frac 1 2 
R^2 (\vp+\bar \vp ) =2R^2 y .
\eeq
The isometry is generated by
\beq
\xi = \frac \partial {\partial x} =-2i \frac \partial {\partial (\vp-\bar \vp )}
\eeq
and the Killing potential
is
\beq\label{8X}
X=R^2 (\vp+\bar \vp ) .
\eeq

The Lie derivative of $k$ is zero, so we are in the simple case 
with $\chi=\vartheta_\alpha =0$.
The T-dual metric is then
\beq
d\hat s^2=\frac 1 {R^2} (d\hat x^2 +d\hat y^2) = \frac 1 {R^2} d \hat zd\bar  {\hat z} ,
\label{realdual}
\eeq
where
\beq\label{8T}
\hat z = \Theta \vert=\hat y+i \hat x ,
\eeq
and  the dual $b$-field is zero. Eq. (\ref{realdual}) looks like the metric that would result from T-dualising on both circles, but to see whether this is the case, we need to be careful with the periodicities. From the T-duality, we know that 
\beq
\hat x\sim \hat x +2\pi~,
\eeq
so the circumference of the $\hat x$ circle is $\frac{2\pi}R$ and we find the dual radius $\hat R=\frac1R$ as expected. 
The 
constraint \re{X8}, together with 
\re{8X} and \re{8T}, gives
\beq
\label{chay}
\hat y=-R^2 y~,
\eeq
so  the periodicity of $\hat y$ is $\hat y \sim \hat y +R^2\tau$.
Using the dual metric \re{realdual}, the circumference of the circle parameterised by $\hat y$ is $R^{-1}R^2 \tau =R \tau$ which is the same as that  of the original circle parameterised by $y$. The T-duality has implemented the change of variables \re{chay} from $y$ to $\hat y$ and this diffeomorphism preserves the circumference of the circle.
Rewriting \re{realdual} in terms of $\hat x$ and the original coordinate $y$, we find
\beq
d\hat s^2=\frac 1 {R^2} d\hat x^2 +R^2d y^2 
\label{realrealdual}
\eeq
which is the result of the standard Buscher rules for T-duality in the $x$-circle.
Thus we see that the $(2,1)$ T-duality, which appears to give a T-duality in two directions, in fact gives a T-duality in just one direction, combined with a diffeomophism whose role is to maintain the extra supersymmetry and the complex geometry.

\section{Geometry and obstructions for $(2,1)$ T-duality} 

\label{Obstruct} 

We start by recalling the results of~\cite{Hull:2006qs} reviewed in section 2. 
For a sigma model with Abelian isometries generated by Killing vectors $\xi_K$, 
the conditions for gauging are that the $u_K$
are globally defined 1-forms that are invariant, $\cl _K u_L=0$, and satisfy 
$\i_K u_L =-\i_L u_K $. 
For T-duality, we require none of these conditions but only that $\i_K \i_L \i_ M H=0$, and 
we introduce a bundle $\hat M$ over $M$ with fibre coordinates $\hat x_K$. 
The metric and $H$-flux are defined by (\ref{abcads}) 
and we take 
\beq 
\hat u_K= u_K+ d\hat x_K . 
\eeq 
We lift the Killing vectors $\xi_K$ on $M$ to Killing vectors 
$\hat \xi_K$ on $\hat M$ satisfying 
(\ref{khatis}) and (\ref{thetis}). The space $\hat M$ can be chosen so that 
$\hat u$ is invariant and globally defined on $\hat M$ with (\ref{hatiu}) satisfied, so that 
the only condition necessary for gauging and hence for T-duality is 
(\ref{ijkH}). The T-dual space is then $\hat M/G$ where $G$ is the Abelian gauge group generated by the $\hat \xi_K$. 

For the $(2,1)$ supersymmetric sigma model to be defined on $M$, $M$ has to be complex with the geometry reviewed in section
3 and the Killing vectors must be holomorphic. Then there are generalised  Killing potentials $Y_K +iX_K$ 
satisfying (\ref{defX}). This can be written as 
\beq 
\xi_{ K} + u_{ K} = dY_K +i(\partial -\bar \partial) X_K , 
\label{defXa} 
\eeq 
with real 1-forms $u_{ K} =u_{ iK} dx^i$, $\xi_{ K} =g_{ij}\xi^j_K dx^i$. 
Locally, we can  absorb $Y_K$ into a   redefinition of $u_K$ as discussed below \re{defX}.
For gauging of the sigma model on $M$ to be possible, the final form of $u_K$ that arises after absorbing all the $dY_K$ terms should be a globally defined one-form; for T-duality, this is not necessary as the $u_K$  do not need to be globally defined.
If the $(2,1)$ sigma-model on $M$
allows a (1,1) gauging\footnote{A (1,1) gauging is the same as the bosonic gauging discussed in section 2.}, then the gauging will be $(2,1)$ supersymmetric 
provided the Killing vectors are holomorphic and the
potentials $X_K$ are globally defined scalars which are invariant: $\cl_KX_L=0$. 
The $(2,1)$ gauging is defined by restricting to the subspace $X=0$ 
and taking a quotient by $G$ to give $X^{-1}(0)/G$. 

For $(2,1)$ T-duality, introducing $n$ extra coordinates $\hat x_K$ ($K=1,\dots, n$) would in general be inconsistent with
supersymmetry; for example, if $n$ is odd, $\hat M$ would have odd dimension and so cannot be complex. 
Instead, we introduce $n$ complex coordinates $\Theta _K$ corresponding to the chiral Lagrange multiplier fields introduced
in section 5. 
This  leads to a complex manifold $\check{M} $ with holomorphic coordinates 
$z^a=(\vp^\alpha, \Theta _K)$ that is a bundle over $M$
with projection $ \check\pi : \check M \to M$ with 
$\check \pi : (\vp^\alpha, \Theta _K) \mapsto \vp^\alpha$. 
A metric $\check g$ and closed 3-form $\check H$ can be chosen on $\check M$ with no $\Theta_K$ components, i.e.\  
\beq \label{abcads1} 
\check g = \check \pi ^* g, \qquad \check H = \check \pi ^* H , 
\eeq 
where 
$\check \pi ^*$ is the pull-back of the projection. 

Writing 
\beq 
\Theta _K =  (\hat y _K +i \hat x_K) , 
\label{thesiss} 
\eeq 
we identify the coordinates $\hat x_K$ with the extra coordinates needed for the (1,1) T-duality. 
Then 
\beq 
\hat u_K= u_K+2d\hat x_K = u_K+i d( \bar\Theta _K - \Theta _K ) 
\eeq 
and we take the Killing vectors on $\check M$ to be the $\hat \xi_K$ given by \re{khatis} and \re{thetis}. 
Now if (\ref{defXa}) holds on $M$ (with $Y_K$ absorbed into $u_K$), then 
on $\check M$ we have 
\beq 
\hat \xi_{ K} + \hat u_{ K} = i(\partial -\bar \partial) X_K + i d(\bar \Theta _K - \Theta _K ) 
\label{defXb} 
\eeq 
and, since $\partial \bar\Theta=\bar\partial\Theta=0$, this can be rewritten as 
\beq 
\hat \xi_{ K} + \hat u_{ K} = +i(\partial -\bar \partial) \check X_K , 
\label{defXc} 
\eeq 
where 
\beq 
\check X_K = X_K + \Theta _K + \bar \Theta _K= X_K +2\hat y_K . 
\label{defXd} 
\eeq 

It follows that $\hat u$ can be chosen so that it is globally defined on $\check M$ and invariant. 
Then $d \check X_K$ will be invariant under the action of the Killing vectors $\hat \xi$, so that 
\beq 
\hat \cl _L \check X_K = C_{LK} 
\label{ffedee} 
\eeq 
for some constants $C_{LK} 
$. Introducing the Killing vectors on $\check M$ 
given by 
\beq 
\check \xi_K= \hat \xi_K - C_{KL} \frac \partial {\partial \hat y_L} , 
\label{eoirtjoijwer} 
\eeq 
we have that the $\check X_K$ are invariant: 
\beq 
\check \cl _L \check X_K = 0 . 
\label{fedo} 
\eeq 
Then the bundle $\check M$ can be defined so that the $\hat u_K$ and $\check X_K$ are globally defined (so that the
transition functions for $u_K$ on $M$ determine those of $\hat x_K$ and 
the transition functions for $X_K$ on $M$ determine those of $\hat y_K$). 
As $\hat u_K$ and $\check X_K$ are globally defined and invariant under the action of $\check \xi$, the isometries generated
by $\check \xi$ can be gauged provided (\ref{ijkH}) holds, giving a $(2,1)$ supersymmetric gauged sigma model on $\check M$.

The gauging imposes the generalised moment map constraints 
\beq 
\check X_K = 0 , 
\label{mommaps} 
\eeq 
which is precisely the condition (\ref{Xthth}) obtained previously. 
This defines a $D+n$ dimensional subspace $\check X^{-1}(0)$ of the $D+2n$ dimensional space $\check M$. 
The gauging then gives the quotient $\check X^{-1}(0)/G$, which is of dimension $D$; this is the T-dual target space. 

\section{Conclusion} 
\label{Summout} 

The $(p,q)$ supersymmetric sigma models (with $p\ge 1,q\ge1$) are a special class of $(1,1)$  models which have extra
geometric structure arising from the $(p-1)+(q-1)$ complex structures of the target space. 
The T-duality of such models can  be analysed using a $(1,1)$ supersymmetric gauging by coupling to $(1,1)$ vector multiplets
and adding $(1,1)$ supersymmetric Lagrange multiplier terms, and this gives the standard Buscher rules when the $b$-field is invariant, and
modified Buscher rules when only $H=db$ is invariant.
However, considerable insight arises from performing T-duality in $(p,q)$ superspace by coupling to $(p,q)$ vector
multiplets and adding $(p,q)$ supersymmetric Lagrange multiplier terms. 
For example, for sigma models with K\" ahler target space, the T-duality is realised as a Legendre 
transformation of the K\"ahler potential \cite{Lindstrom:1983rt,RV}. 

Here we used the $(2,1)$ superspace formulation 
of \cite{DinSeib} and the gauging found in \cite{ AH1,AH2} to find the T-dual $(2,1)$ geometry, realised as a transformation from the potential $k$ to a dual potential $k^D$ that can be viewed as a generalisation of the Legendre transformation of the K\"ahler case. As for the $(2,2)$ case, the supersymmetric gauging  requires   a gauging of a complexification of an isometry group. If the group
$G$ of real isometries to be gauged is generated by Killing vectors $\xi^i_K$, the group which is actually gauged in the
$(2,1)$ supersymmetric gauging is the complexification $G^C$ of $G$  generated by the $\xi^i_K$ together with the vector
fields $J^i{}_j\xi^j_K$. The vector fields $J^i{}_j\xi^j_K$ are in general not Killing vectors, so that they generate
diffeomorphisms that are not isometries. 
This gives a good local picture of the gauging. Global issues are discussed in \cite{Hitchin:1986ea}.
For T-duality with $G$ Abelian, the 
dual metric $g^D$ and the dual torsion potential $b^D$ given by eqs.~(\ref{metdubu})-(\ref{Bdubu}) 
involve a duality in complex coordinates, which does not look like the standard Buscher rules. 
However, as discussed in section 8, just as for the (2,2) supersymmetric case, 
the T-duality transformations in fact correspond to the usual Buscher rules 
for the real isometry group $G$, together with diffeomorphisms that introduce the 
Killing potentials as new coordinates.

The geometry of the T-duality was discussed in sections 2 and 9. The $(1,1)$ T-duality is understood through the construction of a
\lq doubled' manifold $\hat M$ with $n$ extra coordinates $\hat x_K$ arising from the Lagrange multipliers, so that the
coordinates $x^K$ of the torus fibres generated by the $n$ Killing vectors $\xi_K$ are doubled to give a \lq doubled torus'
with $2n$ coordinates $x^K, \hat x_K$. The action of the isometry group $G$ is lifted to $\hat M$, and the T-duality space is
the quotient $\hat M/G$. 

Similarly, the $(2,1)$ duality introduces $n$ extra complex coordinates $\Theta _K$ corresponding to each Killing vector,
giving a space $\check{M} $ with an extra $2n$ real dimensions. The T-dual space is now the symplectic quotient of $\check{M}
$ given by taking the quotient of the subspace $X=0$, giving a generalised moment map and a generalisation of the K\" ahler
quotient construction. 
For generalised K\" ahler spaces, this reduces to the generalised moment map of \cite{LT}, which was constructed from a
$(2,2)$ gauging in \cite{Kapustin:2006ic}. 

\section*{Acknowledgments} 

We are indebted to Rikard von Unge for collaboration in the early stages of this project.
M.A.  wishes to thank Nathan Berkovits, Reimundo Heluani, Kentaro Hori, Nikita Nekrasov and Warren Siegel for
discussions and helpful comments. We thank NORDITA, the GGI, the ICTP-SAIFR, the IMPA, Imperial College London, 
the Mainz ITP, and the SCGP for hospitality while this work was in progress. 
We have benefited from participation in several Simons Summer Workshops at
the SCGP. 
This work was supported by the EPSRC programme grant ``New 
Geometric Structures from String Theory'' EP/K034456/1 and by STFC grant ST/L00044X/1.  
MR was supported in part by NSF grant \# PHY1620628.

\appendix 
\section{Review of chiral and vector representations} 
\label{realrep} 
The superspace constraints \re{21YMdef} can be solved in terms of unconstrained superfields. The chiral representation
solution is given in \re{21nabla-} in terms of $V$ and $A_-$; in this representation, gauge covariant derivatives transform
with the chiral parameter $\La$.  A real representation 
\cite{Wess:1977xn} can be found by writing 
\beq 
e^V=e^{\bar\Om}e^\Om , 
\eeq 
with gauge transformations that depend on both $\La$ and on new, hermitian parameters $K$: 
\beq 
e^{\Om'}=e^{iK}e^\Om e^{-i\La}~~,~~~e^{{\bar\Om}'}=e^{i\bar\La}e^{\bar\Om}e^{-iK}~. 
\eeq 
Note that this is compatible with \re{eV'}.  Performing a similarity transformation 
$\na_\bullet\to e^{\Om}\na_\bullet e^{-\Om}$ on the chiral representation derivatives \re{21nabla-} gives the hermitian
vector representation derivatives 
\ber 
\bar\na_+\!\!&=&\!\!e^\Om\bar D_+ e^{-\Om}~~,~~~\na_+=e^{-\bar\Om} D_+e^{\bar\Om}~,\nn\\ 
\na_-\!\!&=&\!\!e^\Om( D_--iA_-)e^{-\Om}=D_--i(A_--iD_-\Om)+O(\Om^2)~. 
\eer 
We can use the $K$ transformation to make $\Om$ real; in that case, $\Om=\bar\Om=\half V$ and 
we find that 
\beq 
A_--\ihalf D_- V+O(V^2) 
\eeq 
is Hermitian. The higher order terms are absent in the Abelian case (compare \re{realA}). 
\section{Gauge invariance and hermiticity of the action} 
\label{actionprops} 
We now derive the following  useful identities for the operator $\frac{e^L-1}L\equiv\int_0^1 dt\, e^{tL}~$:
\beq\label{elid1} 
\frac{\pa}{\pa V}\left(V\el\right)=e^L~,
\eeq 
which immediately implies the corollary 
\beq 
\label{elid2} 
D_-\left(V\el\right)\equiv (D_-V)\frac{\pa}{\pa V}\left(V\el\right)+V\el D_- = (D_-V)e^L+V\el D_- ~. 
\eeq 
We prove \re{elid1} as an operator relation by applying it to a test function $f(\vp^i)$ where 
$\vp^i$ represents $\vp^\al, \bar\vp^{\bar\al}$: 
\ber 
\frac{\pa}{\pa V}\left(V\el\right)f(\vp^i)&\equiv&\frac\pa{\pa V}\int_0^1dt\,Ve^{tL}f(\vp^i)= 
\frac\pa{\pa V}\int_0^1dt \,Vf(\bar\vp^{\bar0}+tV,\vp^\al,\bar\vp^{\bar\mu})\nn\\ 
&=& 
\int_0^1dt[f(\bar\vp^{\bar0}+tV)+Vtf(\bar\vp^{\bar0}+tV)_{,\bar 0}] 
\nn\\ 
&=& 
\int_0^1dt \frac\pa{\pa t}[t f(\bar\vp^{\bar0}+tV)]=f(\bar\vp^{\bar0}+V) ~, 
\eer 
where the dependence on the spectator fields is suppressed after the first line. 

These identities help us prove several important relations. We start with 
the proof of gauge invariance of \re{Lfin}. 
Its gauge variation may be written as 
\ber\label{gaugevar} 
&&i\La\left((\Lie k_\alpha) D_-\vp^\alpha-(\Lie \bar k_{\bar\alpha})D_-\tilde\vp^{\bar \alpha}\right)-(k_0+\bar k_{\bar
0})D_-\La -D_-\La X-A_-\delta X \nn\\[1mm] 
&&- \delta\left[\left(\el\bar{\mach}_{\bar{\mu}}(\bar\vp)\right) V 
D_-\bar\vp^{\bar\mu}\right]~. 
\eer 
The first term in \re{gaugevar} is 
\ber\label{term1} 
i\La(\Lie k_\alpha) D_-\vp^\alpha=i\La(\vartheta_\al(\vp)+i\chi,_\al(\vp,\tilde\vp))D_-\vp^\al=-\La\chi,_\al
(\vp,\tilde\vp)D_-\vp^\al~, 
\eer 
where a chiral term has been dropped since the measure $\int d^2\theta^+d\theta^-$ annihilates it. 
The second term is 
\ber\nn\label{term2} 
&&-i\bar(\La\Lie \bar k_{\bar\al})
D_-\tilde\vp^{\bar\al}=-i\La(\bar\vartheta_{\bar\al}(\tilde\vp)-i\chi,_{\bar\al}(\vp,\tilde\vp))D_-\tilde\vp^{\bar\al}\\[1mm]
&&= 
-\La\left(i\bar\vartheta_{\bar\mu}(\tilde\vp)D_-\bar\vp^{\bar\mu}+\chi,_{\bar\al}(\vp,\tilde\vp) D_-\tilde\vp^{\bar\al}
\right)~, 
\eer 
where we used that $\bar\vartheta_{\bar 0}=0$, cf.~\re{cthxi0}. 
After partially integrating the $\chi$ terms in \re{term1} and \re{term2} we add them to the third term in~\re{gaugevar}  
and find 
\ber 
-(k_0+\bar k_{\bar 0}-\chi)D_-\La=XD_-\La~, 
\eer 
which cancels the fourth term in \re{gaugevar}. 
The fifth term is zero since $X$ is equivariant. 
The sixth term is 
\ber\label{term6} 
-i(\bar\La-\La)e^L\bar\vartheta_{\bar\mu}(\bar\vp)D_-\bar\varphi^{\bar\mu}+iV\el \bar\vartheta_{\bar\mu},_{\bar
0}(\bar\vp)\bar\La D_-\bar\vp^{\bar\mu}~, 
\eer 
where we used \re{elid1}. The $\La$-term in \re{term6} cancels the $\bar\vartheta$-term in \re{term2}, leaving 
\beq\label{term7} 
-i\bar\La e^L\bar\vartheta_{\bar\mu}(\bar\vp)D_-\bar\varphi^{\bar\mu}+iV\el \bar\vartheta_{\bar\mu},_{\bar 0}(\bar\vp)\bar\La
D_-\bar\vp^{\bar\mu}~. 
\eeq 
From the definition \re{LV} of $L$ we see that 
\ber 
V\bar\vartheta_{\bar\mu},_{\bar 0}=L\bar\vartheta_{\bar\mu}~, 
\eer 
and we have 
\ber 
i\bar\La\left(-e^L+\frac{e^L-1}LL\right)\bar\vartheta_{\bar\mu}(\bar\vp)D_-\bar\varphi^{\bar\mu}=-\left(i\bar\La
\bar\vartheta_{\bar\mu}(\bar\vp)D_-\bar\varphi^{\bar\mu}\right)~, 
\eer 
which is antichiral and again annihilated by the measure. 

As discussed at the end of Section \ref{Gauged21model}, hermiticity of the gauged action follows immediately from the
expression \re{Lgeo}.  We now prove that \re{Lfin} {\em is} \re{Lgeo} (modulo total derivatives). We use the relations
\re{urels}, which we rewrite here for convenience: 
\ber\label{urelsb} 
u_\al&=&i(k_{\al,\bar0}+\bar k_{\bar0,\al}+X_{,\al}) =i(g_{\al\bar0}+X_{,\al}) ~,\\[1mm]\label{urelst} 
\bar\vartheta_{\bar\al}&=&-i(\bar k_{\bar\al,\bar0}-\bar k_{\bar0,\bar\al})-\bar u_{\bar\al}~. 
\eer 
We write out  \re{Lfin}, using the definitions $\tilde\vp=e^L\bar\vp$, $e^L=1+\el L$, and \re{urelst} 
\ber\label{Lfina} 
L_{1g} &= & 
\left[ i\left( k_\alpha D_- \vp^\alpha -\bar k_{\bar{\alpha}}D_- 
\tilde{\vp}^{\bar{\alpha}}\right) - A_{-} X \right]\! (\vp , 
\tilde{\vp}) 
- V\left(\el \bar{\mach}_{\bar{\alpha}}\right)D_-\bar{\vp}^{\bar{\alpha}} \nn\\[1mm] 
&=& 
i( k_\alpha D_- \vp^\alpha -\bar k_{\bar{\alpha}}D_- 
\bar{\vp}^{\bar{\alpha}})(\vp,\bar\vp)-(i\bar k_{\bar0}D_-V+A_-X)(\vp ,\tilde{\vp})\nn\\ 
&&+iV\el ( k_{\alpha,\bar0} D_- \vp^\alpha -\bar k_{\bar{\alpha},\bar0}D_- 
\bar{\vp}^{\bar{\alpha}})- 
V\el(-i(\bar k_{\bar\al,\bar0}-\bar k_{\bar0,\bar\al})-\bar u_{\bar\al})D_-\bar\vp^{\bar\al}\nn\\[1mm] 
&=& 
i( k_\alpha D_- \vp^\alpha -\bar k_{\bar{\alpha}}D_- 
\bar{\vp}^{\bar{\alpha}})(\vp,\bar\vp)-(i\bar k_{\bar0}D_-V+A_-X)(\vp ,\tilde{\vp})\nn\\ 
&&+iV\el ( k_{\alpha,\bar0} D_- \vp^\alpha -\bar k_{\bar0,\bar\al}D_- 
\bar{\vp}^{\bar{\alpha}}) 
+V\el(\bar u_{\bar\al})D_-\bar\vp^{\bar\al}~. 
\eer 
Subtracting \re{Lgeo}, we have: 
\ber\label{Lfindiff} 
0&\stackrel{?}{=}&-iD_-V\bar k_{\bar0}(\vp ,\tilde{\vp})+iV\el ( k_{\alpha,\bar0} D_- \vp^\alpha -\bar k_{\bar0,\bar\al}D_-
\bar{\vp}^{\bar{\alpha}})\nn\\ 
&&-\ihalf D_-VX(\vp ,\tilde{\vp})-V\el\left[(u_\al -\ihalf   X_{,\al})D_-\vp^\al +\ihalf   X_{,\bar\al} 
D_-\bar\vp^{\bar\al}\right]~. 
\eer 
Applying the identity~\re{elid2} on the two $D_-V$ terms in~\re{Lfindiff}, we find, up to total derivatives 
\ber\label{Ldiffsimp} 
0&\stackrel{?}{=}&iV\el D_-\bar k_{\bar0}+iV\el ( k_{\alpha,\bar0} D_- \vp^\alpha - 
\bar k_{\bar0,\bar\al}D_- 
\bar{\vp}^{\bar{\alpha}})\nn\\ 
&&+\ihalf V\el D_-X-V\el\left[(u_\al -\ihalf   X_{,\al})D_-\vp^\al +\ihalf   X_{,\bar\al} 
D_-\bar\vp^{\bar\al}\right]\nn\\[1mm] 
&=&V\el\left[i(\bar k_{\bar0,\al} D_-\vp^\al+\bar k_{\bar0,\bar\al} D_-\bar\vp^{\bar\al}) 
+i( k_{\alpha,\bar0} D_- \vp^\alpha -\bar k_{\bar0,\bar\al}D_-\bar{\vp}^{\bar{\alpha}}) 
\right]\nn\\ 
&&+V\el\left[\ihalf (X_{,\al}D_-\vp^\al+X_{,\bar\al}D_-\bar\vp^{\bar\al})-(u_\al -\ihalf   X_{,\al})D_-\vp^\al -\ihalf  
X_{,\bar\al} 
D_-\bar\vp^{\bar\al} 
\right]\nn\\[1mm] 
&=&V\el\left[i(\bar k_{\bar0,\al} +\bar k_{\al,\bar0}) D_-\vp^\al+(iX_{,\al}-u_\al)D_-\vp^\al\right]~. 
\eer 
This vanishes because of \re{urelsb}. 

If we keep all the total derivative terms, we find that the difference between the two Lagrangians 
\re{Lfin} and \re{Lgeo} is the total derivative 
\ber 
L_{1g}-L'_{1g}=-\ihalf D_-\left(V\frac{e^L -1}{L}Z(\vp,\bar\vp)\right), 
\eer 
where $Z\equiv X+2\bar k_{\bar 0}=-k_0+\bar k_{\bar 0}+\chi $ is defined in (\ref{Zdef}-\ref{KG}). 

\section{Calculation of $A_-(\vp)$}\label{derAm} 
As observed in 
section \ref{21Lfindual}, the spinor connection $A_-$ enters the action that we use for T-duality--the sum of \re{Lgeo} and 
\re{Lagrange2K}--as a Lagrange multiplier, and hence is not needed; for completeness we present its calculation here. 

The expression for the potential $A_-$ is found from the $V$ field equation. 
The variation of $V$ in the sum of \re{Lgeo} and  \re{Lagrange2K} is 
\ber\nn 
&&\delta V\frac {\partial}{\partial V }(L_{1g}+L_{\Theta})\\[1mm]\nn 
&&=\delta V\frac {\partial}{\partial V }\left( 
V\frac{e^L-1} 
L\left[ 
(u_\alpha-\ihalf X,_\alpha)D_-\vp^\al+(\bar u_{\bar\alpha}+\ihalf X,_{\bar\alpha})D_-\bar\vp^{\bar\al} 
\right]\right.\\[1mm]\nn 
&&~~~~~~~~~ ~~~~~-(A_--\ihalf D_-V)X(\vp,\tilde \vp)-iVD_-\bar\Theta\bigg)\\[1mm]\nn 
&&=\delta V\bigg(e^L \left[ 
(u_\alpha-\ihalf X,_\alpha)D_-\vp^\al+(\bar u_{\bar\alpha}+\ihalf X,_{\bar\alpha})D_-\bar\vp^{\bar\al} 
\right]\\[1mm]\nn 
&&~~~~~~~~~~-(A_--\ihalf D_-V)X_{\bar 0}(\vp,\tilde \vp)-iD_-\bar\Theta\bigg)+\ihalf  XD_-\delta V\\[1mm] 
&&=\delta V\bigg(u_\alpha D_-\vp^\al+\bar u_{\bar\alpha}D_-\bar\vp^{\bar\al}-i X,_\alpha D_-\vp^\al-A_-X,_{\bar
0}-iD_-\bar\Theta\bigg)~, 
\eer 
where $u, \bar u$ and $X$ now all depend on $\vp$ and $\tilde\vp$. The $V$ field equation results from setting the expression
multiplying $\delta V$ to zero, and determines $A_-$ to be 
\ber\label{A} 
A_-=\frac 1 {X,_{\bar 0}}\bigg(u_\alpha D_-\vp^\al+\bar u_{\bar\alpha}D_-\bar\vp^{\bar\al}-i X,_\alpha
D_-\vp^\al-iD_-\bar\Theta\bigg) . 
\eer 
We note that, using the $A_-$ field equation \re{Aeq}, we have 
\ber\nn\label{Ac} 
A_--iD_-V&\!\!=\!\!&\frac 1 {X,_{\bar 0}}\bigg(u_\alpha D_-\vp^\al+\bar u_{\bar\alpha}D_-\bar\vp^{\bar\al}-i X,_\alpha
D_-\vp^\al-iD_-\bar\Theta\bigg)\\[1mm]\nn 
&&\qquad+\frac i {X,_{\bar 0}}\bigg( X,_\alpha D_-\vp^\al+X,_{\bar\al}
D_-\bar\vp^{\bar\al}+D_-\Theta+D_-\bar\Theta\bigg)\\[1mm] 
&\!\!=\!\!&\frac 1 {X,_{\bar 0}}\bigg(u_\alpha D_-\vp^\al+\bar u_{\bar\al}D_-\bar\vp^{\bar\al}+i X,_{\bar\al}
D_-\bar\vp^{\bar\al}+iD_-\Theta\bigg)~. 
\eer 
Since X is real and equivariant, $X_0=X_{\bar0}$, and \re{Ac} is indeed the complex conjugate of \re{A}. Hence the
combination 
$A_--\ihalf D_-V$ in \re{realA}, which is the average of \re{A} and \re{Ac}, is manifestly real. 

\section{Reduction} 
\label{redux} 
In this appendix we reduce $(2,2)$ models to $(2,1)$ superspace and $(2,1)$ models to $(1,1)$ superspace. 

\subsection{Reduction of a K\"ahler $(2,2)$ sigma model  to $(2,1)$ superspace} 

The gauged $(2,2)$ action for chiral superfields $\phi$ reads \cite{Hull:1985pq} 
\ber 
\int d^2xd^2\theta d^2\bar\theta\left(K(\phi, \bar\phi)-\half V\frac{e^L-1}LX 
\right) ,
\eer 
where $-\half X$ is the Killing potential for the isometry.  To reduce to $(2,1)$ we write 
\ber 
\bbD{-}=D_--iQ_- ~~~\Rightarrow~~~ D_-=\half( \bbD{-} + \bbDB- )~~,~~ 
Q_-=\ihalf( \bbD{-} - \bbDB-  )~, 
\eer 
and act with $Q_-$ on $K$, splitting the fermionic measure as follows:
\ber 
d^2\theta d^2\bar\theta\sim \bbD{+} \bbDB{+} D_-Q_-~. 
\eer 
We also write the gauge covariant derivative as 
\beq\label{d1Am} 
\na_-\equiv D_--iA_-=\half(\bbnabla_-+\bar\bbnabla_-)_|\equiv 
\half( \bbD{-} +\bbD- V+ \bbDB-)_|=D_-+\half(D_-V-iQ_-V)_| ~~,
\eeq 
which implies 
\beq 
Q_-V_|=2(A_--\ihalf D_-V) \ ,
\eeq 
where a vertical bar denotes the reduction to $(2,1)$ superspace.  Note that this is proportional to the real combination
\re{realA}. 
We use the chiral constaint 
\beq 
\bbDB{-}\phi=0~~\Rightarrow ~~Q_-\phi_|=iD_-\phi~, 
\eeq 
and keep the $(2,2)$ notation ($\phi,V$) for the $(2,1)$ superfields 
(that is, we don't bother writing $\vp:=\phi_|$, etc.). 

This leads to the $(2,1)$ action 
\ber\nn 
&&\int d^2x\bbD{+} \bbDB{+} D_-\bigg(i(K_{,\al} D_-\phi^\al-K_{,\bar\al} D_-\bar\phi^{\bar\al})\\[1mm] 
&&-2(A_--\ihalf D_-V)\half X 
-V\frac{e^L-1}L\ihalf(X_{,\al} D_-\phi^\al-X_{,\bar\al} D_-\bar\phi^{\bar\al})  
\bigg) .
\eer 
As expected, this is the $(2,1)$ action \re{Lgeo}  with $k_\al=K_{,\al}$ and 
$u_\al=0$ because the geometry is K\"ahler. 

\subsection{Reduction of a general $(2,1)$ sigma model  to $(1,1)$ superspace} 
The general gauged  $(2,1)$ model \re{Lgeo} is reduced to  $(1,1)$ superspace using similar techniques to those in the
previous subsection.  However, the reduction of the gauge multiplet is somewhat different. 

As above, we define 
\ber 
\bbD{+}=D_+-iQ_+
\eer 
and write the measure as 
\ber 
\bbD{+} \bbDB{+} D_-\sim D_+D_-Q_+~. 
\eer 
We can define gauge covariant $(1,1)$ derivatives from the $(2,1)$ derivatives in section \ref{21YMmultiplet}; however, since
we need to distinguish them, we will write $\bbnabla$ for the $(2,1)$ gauge covariant derivatives and $\na$ for the  $(1,1)$
derivatives. In $(1,1)$ superspace, the natural group has a real superfield gauge parameter, and does {\em not} involve a
complexification. Instead, the real part of the $(2,1)$ complex gauge parameter $\La$ is used to gauge away $V_|$. This is the
$(1,1)$ version of Wess-Zumino gauge; note that 
$V_|=D_\pm V_|=0$, but $(Q_+V)_|\ne0$. 
In Wess-Zumino gauge, the $(1,1)$ objects that 
we find are independent of whether we are in chiral, anti-chiral, or real representation. 

Similarly to \re{d1Am}, we define the $(1,1)$ gauge covariant derivative: 
\beq\label{d2Ap} 
\na_+\equiv D_+-iA_+=\half(\bbnabla_++\bar\bbnabla_+)_|\equiv 
\half( \bbD{+} +\bbD+ V+ \bbDB+)_|=D_++\half(D_+V-iQ_+V)_|~, 
\eeq 
where a vertical bar denotes the reduction to $(1,1)$ superspace.  Here we have used \re{21nabla-}. 
Because in Wess-Zumino gauge $(D_+V)_|=0$, we find 
\beq 
Q_+V_|=2A_+~. 
\eeq 
We also reduce the field strengths $W$, $\bar W$ in \re{Wmin21},\re{Wbarmin21}: 
\beq 
W=-i\bbDB+A_-~~,~~~\bar W= -i\bbD+(A_--iD_-V)~. 
\eeq 
Since in Wess-Zumino gauge $(D_-V)_|=0$, we find 
\beq 
W_|\equiv (-iD_+A_-+Q_+A_-)_| ~~,~~~ \bar W_|\equiv (-iD_+A_--Q_+A_-)_|-2iD_-A_+~. 
\eeq 
Then the real part of $W$ is just the $(1,1)$ gauge field strength: 
\beq 
\half(W+\bar W)_|=-i(D_+A_-+D_-A_+)=\{\na_+,\na_-\}~, 
\eeq 
and the imaginary part is a real scalar field: 
\beq 
\ihalf(\bar W-W)_|=(-iQ_+A_-)_|+D_-A_+=: -\half s~~\Rightarrow~~ (Q_+A_-)_|=-i(\half s+D_-A_+)~. 
\eeq 
We will use the following identity below: 
\beq 
Q_+(A_--\ihalf D_- V)_|=-\ihalf s~. 
\eeq 

We now push in the generator of the nonmanifest left supersymmetry $Q_+$: 
\ber\label{Lgeo2d1}\nn 
&&\int d^2xd^2\theta^+d\theta^-L_{1g} \\[1mm] 
&&=\int d^2x D_+D_-Q_+ 
\bigg( i\left( k_\alpha D_- 
\vp^\alpha -\bar k_{\bar{\alpha}}D_- 
\bar{\vp}^{\bar{\alpha}} \right) \! (\vp , \bar\vp) 
- (A_--\ihalf   D_-V) X(\vp ,\tilde{\vp})\nn\\[1mm]\nn 
&&~~~~~~~~~~~~~~~~~+\,V\frac{e^L -1}{L} 
\left[(u_\al -\ihalf   X_{,\al})D_-\vp^\al +(\bar u_{\bar\al}+\ihalf  
X_{,\bar\al})D_-\bar\vp^{\bar\al}\right]\bigg)\\[1mm]\nn 
&&=\int d^2x D_+D_-\bigg(2k_\al,_{\bar\beta}D_+\bar\vp^{\bar\beta}D_-\vp^\al+2\bar
k_{\bar\al},_{\beta}D_+\vp^{\beta}D_-\bar\vp^{\bar\al}+\ihalf sX\\[1mm]\nn 
&&~~~~~~~~~~~~~~~~~+A_-\left[iX,_\al D_+\vp^\al-iX,_{\bar\al}D_+\bar\vp^{\bar\al}+2X,_0A_+\right]\\[1mm] 
&&~~~~~~~~~~~~~~~~~+ 2A_+\left[v_\al D_-\vp^\al+\bar v_{\bar\al}D_-\bar\vp^{\bar\al} 
\right]\bigg)~, 
\eer 
where we have integrated by parts to eliminate the $D_+D_-\vp$ and $ D_+D_-\bar\vp$ terms, and use the shorthand notation
\ber 
\label{D18}
v_\al = u_\al-\ihalf X,_\al~. 
\eer 
This is not yet manifestly left-right symmetric. We use \re{urels}: 
\ber 
u_\al=i(k_\al,_{\bar 0}+\bar k_{\bar 0},_\al )+iX,_\al=i(g_{\al\bar0}+X,_\al) 
\eer 
to rewrite this as 
\ber\label{Lgeo2d2}\nn 
&& \int d^2x D_+D_-\bigg(2k_\al,_{\bar\beta}D_+\bar\vp^{\bar\beta}D_-\vp^\al+2\bar
k_{\bar\al},_{\beta}D_+\vp^{\beta}D_-\bar\vp^{\bar\al}+\ihalf sX+2X,_0A_-A_+\\[1mm]\nn 
&&~~~~~~~~~~~~~~~~~+A_-\left[(u_\al -ig_{\al\bar0})D_+\vp^\al+(\bar
u_{\bar\al}+ig_{0\bar\al})D_+\bar\vp^{\bar\al}\right]\\[1mm] 
&&~~~~~~~~~~~~~~~~~+ A_+\left[(u_\al +ig_{\al\bar0})D_-\vp^\al+(\bar
u_{\bar\al}-ig_{0\bar\al})D_-\bar\vp^{\bar\al}\right]\bigg) ~.
\eer 
Using the definition of gauge covariant derivatives: 
\ber\nn 
&&\na_\pm \vp^0=D_\pm \vp^0-\xi^0A_\pm=D_\pm \vp^0-iA_\pm\\[1mm] 
&&\na_\pm \bar\vp^{\bar0}=D_\pm \bar\vp^{\bar0}-\xi^{\bar0}A_\pm=D_\pm \bar\vp^{\bar0}+iA_\pm 
\eer 
as well as \re{geometry} in the form 
\beq 
2k_\al,_{\bar\beta}=g_{\al\bar\beta}-b_{\al\bar\beta}~~,~~~ 
2k_{\bar\beta},_\al=g_{\al\bar\beta}+b_{\al\bar\beta}~, 
\eeq 
we  rewrite \re{Lgeo2d2} as 
\ber\label{LgaugeD22} \nn
&&\int d^2x D_+D_-\bigg(g_{\al\bar\beta}\left[\na_+\bar\vp^{\bar\beta}\na_-\vp^\al+ 
\na_+\vp^{\al}\na_-\bar\vp^{\bar\beta}\right]+ 
b_{\al\bar\beta}\left[D_+\vp^{\al}D_-\bar\vp^{\bar\beta}- 
D_+\bar\vp^{\bar\beta}D_-\vp^\al\right]\\[1mm]\nn 
&&~~~~~~~~~~~~~~~~~+\ihalf sX+A_-\left[u_\al D_+\vp^\al 
+\bar u_{\bar\al}D_+\bar\vp^{\bar\al}\right]+ 
A_+\left[u_\al D_-\vp^\al+\bar u_{\bar\al}D_-\bar\vp^{\bar\al}\right]\bigg)~. \\
\eer 
Except for the $sX$ term discussed below, this is precisely the general gauged sigma model 
\re{Sggauge},\re{wzb2} in $(1,1)$ superspace. 

Next, we reduce the Lagrange multiplier term $L_\Theta$ \re{Lagrange2K} to $(1,1)$ superspace: 
\ber\nn\label{LlagD23}
L_\Theta &\!\!=\!\!& -\int d^2x D_+D_-Q_+\left[( \Theta + \bar{\Theta} ) A_- 
+iD_- \bar{\Theta} V \right]\\[1mm] 
&\!\!=\!\!& 
-\int d^2x D_+D_-Q_+\left[( \Theta + \bar{\Theta} ) (A_- -\ihalf D_-V) 
+\ihalf D_-(\bar{\Theta}-\Theta) V \right]\nn\\[1mm] 
&\!\!=\!\!& 
-\int d^2x D_+D_-\left[iD_+(\Theta-\bar\Theta)A_--\ihalf s ( \Theta + \bar{\Theta} ) 
+iA_+D_-(\bar{\Theta}-\Theta)\right]\nn\\[1mm] 
&\!\!=\!\!& 
\int d^2x D_+D_-\left[-i(A_-D_++A_+D_-)(\bar\Theta-\Theta)+\ihalf s ( \Theta + \bar{\Theta} )\right]~.  
\eer 
Combining \re{LgaugeD22} and \re{LlagD23}, we find the usual real T-duality with in addition the $(1,1)$ superfield $s$ acting
as a Lagrange multiplier to impose the  condition 
\beq 
 ( \Theta + \bar{\Theta} ) +X(\vp,\bar\vp)=0~. 
 \eeq 
This is a diffeomorphism that expresses $ \Theta + \bar{\Theta}$ in terms of $\vp,\bar\vp$; because of the 
isometry in the original model,  $ \Theta + \bar{\Theta}$  does {\em not} depend on $x^0:=\ihalf(\bar\vp^{\bar0}-\vp^0)$, which is
the coordinate dual to $\hat x^0:=\ihalf(\bar{\Theta}-\Theta)$.

\end{document}